\documentclass[fleqn,usenatbib]{mnras}

\usepackage{newtxtext,newtxmath}

\usepackage[T1]{fontenc}

\DeclareRobustCommand{\VAN}[3]{#2}
\let\VANthebibliography\thebibliography
\def\thebibliography{\DeclareRobustCommand{\VAN}[3]{##3}\VANthebibliography}

\usepackage{graphicx}	
\usepackage{amsmath}	
\usepackage{booktabs}
\usepackage{tablefootnote}
\usepackage{gensymb}
\usepackage{comment}
\usepackage{etoolbox}

\newcommand{\orcid}[1]{%
  \ifblank{#1}{}{%
    \href{https://orcid.org/#1}{\includegraphics[width=10pt]{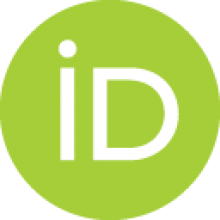}}%
  }%
}

\newcommand{\citnp}[1]{\textcolor{blue}{#1} (\textcolor{blue}{in prep})}

\newcommand{\mdot}{$\dot{M}$}
\newcommand{\mstar}{$M_*$}
\newcommand{\mmol}{$M_{\rm mol}$}
\newcommand{\mion}{$M_{\rm ion}$}
\newcommand{\pdot}{$\dot{p}$}
\newcommand{\edot}{$\dot{E}$}

\newcommand{\vlos}{$\upsilon_{\rm LOS}$}
\newcommand{\etam}{$\eta_M$}
\newcommand{\etae}{$\eta_E$}
\newcommand{\etap}{$\eta_p$}

\title[Resolved, multiphase outflow in ESO~484-036]{The GECKOS survey: Resolving the molecular and ionised gas in the galactic outflow of ESO~484-036}

\author[J. Hernández-Yévenes et al.]{
J. Hernández-Yévenes\ \orcid{0000-0001-5845-7538}$^{1,}$\thanks{Contact e-mail: \href{mailto:jheryev@gmail.com}{jheryev@gmail.com}},
D. B. Fisher\ \orcid{0000-0003-0645-5260}$^{1,2}$,
B. Mazzilli Ciraulo\ \orcid{}$^{1,2}$,
R. L. Davies\ \orcid{0000-0002-3324-4824}$^{1,2}$,
M. Martig\ \orcid{}$^{3}$,
\newauthor
A. Fraser-McKelvie\ \orcid{0000-0001-9557-5648}$^{4,2}$,
J. van de Sande\ \orcid{}$^{5,2}$,
M. R. Hayden\ \orcid{}$^{6}$,
R. Elliot\ \orcid{}$^{1,2}$,
E. Emsellem\ \orcid{}$^{6}$,
F. Combes\ \orcid{0000-0003-2658-7893}$^{7}$,
\newauthor
A. D. Bolatto\ \orcid{}$^{8}$,
J. Bland-Hawthorn\ \orcid{}$^{9,2}$,
L. Cortese\ \orcid{}$^{10,2}$,
T. A. Davis\ \orcid{}$^{11}$,
B. Catinella\ \orcid{}$^{10,2}$,
L. M. Valenzuela\ \orcid{0000-0002-7972-9675}$^{12}$,
\newauthor
S. M. Croom\ \orcid{0000-0003-2880-9197}$^{10}$,
S. A. Fortuné\ \orcid{}$^{12}$,
L. A. Silva-Lima\ \orcid{0000-0003-3490-9063}$^{13}$,
C. López-Cobá\ \orcid{0000-0003-1045-0702}$^{6}$,
A. Mailvaganam\ \orcid{}$^{14,15,2}$
\newauthor
and G. van de Ven\ \orcid{}$^{16}$
\\
$^{1}$Centre for Astrophysics and Supercomputing, Swinburne University of Technology, John Street, Hawthorn, 3122, Australia\\
$^{2}$ARC Centre of Excellence for All Sky Astrophysics in 3 Dimensions (ASTRO 3D), Australia\\
$^{3}$Astrophysics Research Institute, Liverpool John Moores University, 146 Brownlow Hill, Liverpool L3 5RF, UK\\
$^{4}$European Southern Observatory, Karl-Schwarzschild-Straße 2, Garching, 85748, Germany\\
$^{5}$School of Physics, University of New South Wales, Sydney, NSW 2052, Australia\\
$^{6}$Homer L. Dodge Department of Physics \& Astronomy, University of Oklahoma, 440 W. Brooks St., Norman, OK 73019, USA\\
$^{7}$Observatoire de Paris, LUX, Coll\`ege de France, CNRS, PSL, Sorbonne University, 75014, Paris, France\\
$^{8}$Department of Astronomy, University of Maryland, College Park, MD 20742, USA\\
$^{9}$Sydney Institute for Astronomy, School of Physics, A28, The University of Sydney, NSW, 2006, Australia\\
$^{10}$International Centre for Radio Astronomy Research (ICRAR), The University of Western Australia, 35 Stirling Hwy, Crawley, WA 6009, Australia\\
$^{11}$Cardiff Hub for Astrophysics Research \& Technology, School of Physics \& Astronomy, Cardiff University, Queens Buildings, Cardiff, CF24 3AA, UK\\
$^{12}$Universitäts-Sternwarte, Fakultät für Physik, Ludwig-Maximilians-Universität München, Scheinerstr. 1, 81679 München, Germany\\
$^{13}$Núcleo de Astrofísica, Universidade Cidade de São Paulo, Rua Galvão Bueno, 868, São Paulo, Brazil\\
$^{14}$School of Mathematical and Physical Sciences, Macquarie University, NSW 2109, Australia\\
$^{15}$Macquarie University Astrophysics and Space Technologies Research Centre, Sydney, NSW 2109, Australia\\
$^{16}$Department of Astrophysics, University of Vienna, Türkenschanzstraße 17, 1180 Vienna, Austria
}
\date{Accepted XXX. Received YYY; in original form ZZZ}

\pubyear{2026}

\begin{document}
\label{firstpage}
\pagerange{\pageref{firstpage}--\pageref{lastpage}}
\maketitle

\begin{abstract}
We present a spatially resolved, multiphase study of the outflow in the edge-on starburst galaxy ESO~484-036 from the GECKOS survey, combining VLT/MUSE H$\alpha$ and ALMA CO(1$-$0) observations to analyse the atomic ionised and cold molecular gas. Both show extraplanar emission consistent with a conical outflow. Ionised gas is enclosed by molecular gas, which is detected up to 2.5 kpc from the disc. Molecular gas dominates near the disc, except at the nuclear base, while ionised gas extends beyond 3 kpc. The deprojected outflow velocities are $\lesssim400\ \rm km\ s^{-1}$ in both phases and are consistent with ballistic motion, with some gas possibly falling back onto the disc. We find that the mass outflow rates are in the range of $\dot M_{\rm ion}\sim1-5\ \rm M_\odot\ \rm yr^{-1}$ and $\dot M_{\rm mol}\sim13-54\ \rm M_\odot\ \rm yr^{-1}$, giving mass loading factors of $\eta_{M\rm, ion}\sim 0.1-0.6$ and $\eta_{M\rm, mol}\sim 1.5-6.2$. These ranges reflect velocity and geometric uncertainties. Despite the short depletion time ($\tau_{\rm dep} = 16-48\rm\ Myr$), the outflow may regulate rather than permanently quench the gas reservoir. Energy loading ($\eta_E\leq0.16$) and momentum loading ($\eta_p\lesssim1$) support a purely starburst-driven outflow. Comparing ESO~484-036 with a literature sample, we find a systematic 1~dex shift in mass-loading relations when molecular gas is included. This produces a $\sim3.5$~dex discrepancy with cosmological simulations in $\eta_{M\rm, mol}/\eta_{M\rm, ion}$, implying that current models strongly underpredict cold gas production and the role of short-range recycling flows in starburst galaxies.
\end{abstract}



\section{Introduction}

Galactic outflows are ubiquitous, high-velocity, multiphase gas streams ejected from high star forming galaxies \citep{Veilleux2005, Carniani2016, Veilleux2020}. As crucial components of the baryon cycle, these winds regulate galaxy evolution, shaping the galaxy mass function and stellar properties \citep{Chevalier1985, Thompson2024}. Outflows primarily suppress star formation (SF) through two mechanisms: ejective feedback, which removes gaseous fuel, and preventive or delayed feedback, which heats the surrounding medium to prevent new gas accretion \citep{Somerville2015, Fluetsch2019, Thompson2024}. They are also highly effective at transporting heavy elements and dust, enriching the circumgalactic medium (CGM) and, in extreme cases, the intergalactic medium (IGM) \citep{Christensen2018, Cameron2021, Hamel2024}.

Starbursts and active galactic nuclei (AGN) are the primary drivers of galactic outflows \citep[e.g.,][]{Bolatto2013, Herrera2020}. In starbursts, highly concentrated star formation launches superwinds powered by stellar feedback, specifically energy and momentum injection from clustered supernovae (SNe) and massive stellar ejecta \citep{Veilleux2005, Fielding2018, Xu2022}. This injection creates the hot volume-filling phase ($T\gtrsim10^6$ K), surrounded by warm ionised material ($T\gtrsim10^4$ K), which together carry the bulk of the outflow energy \citep{Strickland2009, Shopbell1998}. The ram pressure from this hot flow accelerates and entrains colder molecular and atomic gas clouds \citep{Leroy2015, Martini2018}. This cold phase ($T\lesssim 10^2$ K) surrounds the hotter components and carries the bulk of the outflow mass \citep{Fluetsch2019, Bolatto2024}. Additional mechanisms, including radiation pressure on dusty clouds \citep[e.g.,][]{Barcos2018} and cosmic rays \citep[e.g.,][]{Socrates2008, Sike2025}, also contribute to driving starburst winds, which typically yield mass outflow rates comparable to or exceeding the star formation rate (SFR).

Although outflows have been observed in hundreds of galaxies across a wide range of wavelengths \citep[e.g.,][]{Rupke2005, Rubin2014,Roberts2019,Forster2019, Herrera2025}, spatially resolved observations exist for a handful of cases \citep[e.g.,][]{Cicone2014, Bolatto2021}, and multiphase-resolved observations for even fewer systems, including the well-known starburst galaxy M82 \citep[e.g.,][]{Shopbell1998,Lopez2020,Krieger2021,Fisher2025}, NGC~253 \citep[e.g.,][]{Bolatto2013,Krieger2019, Cronin2025} and the late-stage merger ultra-luminous infrared galaxy (ULIRG), Arp 220 \citep[e.g.,][]{Barcos2018, Perna2020, Chandar2023}. Resolved observations of outflows predominantly driven by star formation are even rarer in edge-on systems \citep[e.g.,][]{McPherson2023}. 

Molecular gas, present in the cold phase, fuels star formation and represents the main mass component of outflows in the best studied cases \citep{Leroy2015,Fluetsch2019}. Although its removal directly impacts future star formation, models struggle to explain its presence in outflows \citep{Kim2018, Rathjen2023}. Hotter phases are expected to destroy the cold gas before it reaches the observed high velocities \citep{Gronke2018, Richie2024}. Proposed survival mechanisms include turbulent radiative mixing layers \citep{Fielding2022}, consistent with M82 observations \citep{Fisher2025}, and thermal instabilities \citep{Thompson2016}. Molecuar gas velocities indicate that it likely remains within the halo, ascending into the inner CGM, and eventually fuelling future star formation. Demonstrating the existence and survival of this cold component is therefore crucial. Millimeter-wave observations use Carbon Monoxide (CO) transitions, primarily the $J=1-0$ ground rotational transition, as the fundamental tool to survey the structure and kinematics of cool molecular gas in galaxies and outflows \citep{Bolatto2013review}. Since CO possesses a low excitation energy and critical density, it easily excites even in cold molecular clouds, making it the workhorse tracer for the bulk distribution of molecular hydrogen ($\rm H_2$). 

Ionised gas, present in the warm phase, is a ubiquitous component of galactic outflows \citep{Veilleux2003, Xu2023, Cronin2025}. Observations of some targets indicate that the ionised gas represents a much smaller fraction of the total mass compared to the cooler atomic or molecular phases in luminous galaxies \citep{Fluetsch2019, Rupke2019,  Mazzilli2025}. This phase is key for understanding the kinematics and energetics of the wind, as it is thought to trace material either photoionised by the central source or entrained and shocked by the faster, hotter wind fluid \citep{Richie2024,Lopez2025hst}. This gas is studied using various emission and absorption line tracers, including emission lines such as H$\beta$, [OIII], [NII], and [SII], and absorption lines from photoionised metals such as Si IV. The Balmer recombination line H$\alpha$ ($\lambda6563$ \AA) is one of the primary optical diagnostics used, as its dust-corrected luminosity ($L_{\rm H\alpha}$) provides a measure of the total ionised gas content.

Current facilities such as ALMA and JWST offer the high spatial resolution needed to resolve this cold phase and study the molecular gas using different CO transitions \citep[e.g.,][]{Bolatto2013} and its dust content using Polycyclic Aromatic Hydrocarbons \citep[PAHs; e.g.,][]{Fisher2025}, the latter of which may be crucial to explain the existence of the cold phase. Combining these with observations of the ionised gas, with tracers such as H$\alpha$ or [OIII] using VLT Multi Unit Spectroscopic Explorer (MUSE), reveals how hotter and more energetic phases ionise the gas, thus showing the interaction with the cold gas \citep[][]{Leroy2015}.

The mass outflow rate (\mdot) is a key physical parameter of galactic outflows, measuring the quantity of gas expelled from a galaxy per unit time. Although the outflow velocity ($\upsilon_{\rm out}$) determines the fate of the gas, distinguishing between material that escapes the halo and that which is recycled, \mdot\ defines the immediate impact of feedback on the host galaxy. A higher \mdot\ removes material from the ISM more rapidly, effectively cutting off the fuel for future star formation and limiting stellar mass growth. By controlling how fast gas is expelled, \mdot\ separates winds that simply recycle material from those that can clear the galaxy of gas and fully quench star formation.

The ratio between the mass outflow rates of the molecular ($\dot{M}_{\rm mol}$) and ionised ($\dot{M}_{\rm ion}$) gas phases is usually used as a diagnostic tool to constrain the primary energy source driving the wind. Starburst-driven winds often exhibit comparable mass-loss rates between these two phases ($\dot{M}_{\rm mol}/\dot{M}_{\rm ion} \approx 1$), while AGN dramatically boost the molecular mass flux ($\dot{M}_{\rm mol}/\dot{M}_{\rm ion} \gg 1$), generating molecular rates up to two or three orders of magnitude higher than the ionised phase \citep{Fluetsch2019, Roberts2019}. However, cases in the literature such as NGC~253 \citep{Krieger2019, Cronin2025} and NGC~1482 \citep{Salak2020} show that pure starbursts can also present high $\dot{M}_{\rm mol}/\dot{M}_{\rm ion}$ without the need of an AGN component, challenging the use of this diagnostic.

In this work, we present a higher resolution detection of the ionised outflow reported in \cite{Veilleux2003} in ESO~484-036, the molecular outflow detected using CO(1$-$0), and discuss how both phases relate in this galaxy. In Sect.~\ref{sec:data} we present our target, ESO~484-036 and the data used in this work, their specifications and reduction processes, in Sect.~\ref{sect:describing_obs} we present and describe the extraplanar emission detected with both ionised and molecular gas, in Sect.~\ref{sec:geometry} we present how the geometry of the extraplanar emission is constrained, in Sect.~\ref{sec:dynamics} we present our approach to calculate the dynamics of this extraplanar emission, in Sect.~\ref{sec:outrates} we present and discuss the main physical properties of the outflow causing the extraplanar emission and compare it with other observational and theoretical results in the literature, and in Sect.~\ref{sec:conclusion} we summarise our work.

In this paper, we adopt a distance of 68.7 Mpc \citep{Fraser2025}, such that 1$^{\prime\prime}$ corresponds to 0.33 kpc.

\section{Observations}\label{sec:data}

\subsection{Our target: ESO~484-036}\label{sec:target}

\begin{figure*}
    \centering
    \includegraphics[width=\textwidth]{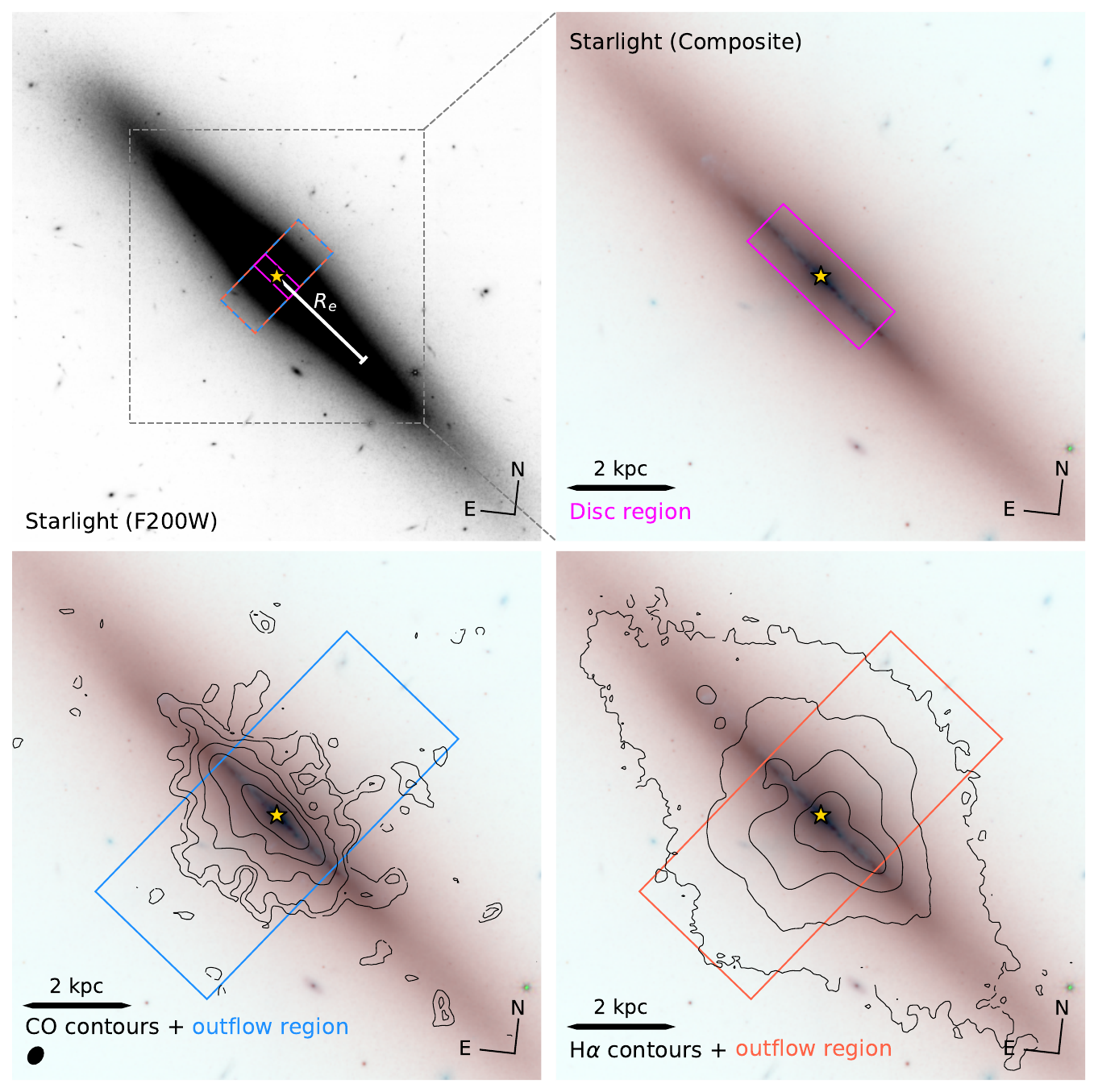}
    \caption{ESO~484-036 starlight image showcasing the multiphase outflow structure. All panels show starlight in log scale. Top Left: $55^{\prime\prime} \times 55^{\prime\prime}$ JWST F200W image with a different stretch to show the complete extent of the starlight disc. Half-light radius ($R_e=4.2\rm\ kpc$) is shown in white. Outflow (red/blue) and disc (magenta) regions are presented. Remaining three panels focus on the central region showed in the grey square ($30^{\prime\prime} \times 30^{\prime\prime}$), displaying a JWST composite image (color-inverted) combining F430M (red), F200W (green), and F150W (blue). Top Right: Shows the galactic disc; the magenta rectangle is defined to avoid this disc region for outflow calculations. Bottom Left: Shows ALMA CO(1$-$0) contours (S/N$=3.5$), with the blue rectangle defining the outflow region. Bottom Right: Shows VLT/MUSE H$\alpha$ contours (S/N$=5$, used for science), where the red rectangle defines the outflow region. The golden star marks the galactic centre (used for major/minor axes definition). Both CO and H$\alpha$ contours show extraplanar gas, a clear probe for the presence of a multiphase outflow. Bottom right panel also displays the cross-like structure described by \protect\cite{Veilleux2003}.}
    \label{fig:RGB}
\end{figure*}

ESO~484-036 (Fig.~\ref{fig:RGB}) is a nearby (68.7 Mpc) edge-on ($i\sim87.5^\circ$, \textcolor{blue}{van de Sande et al. in prep.}) starburst galaxy with a stellar component that exhibits a complex structure, including a stellar bar and a boxy-peanut bulge \citep{Fraser2025}. It has a redshift of $z=0.017$, a recession velocity of $5058\rm\ km\ s^{-1}$ \citep{Makarov2014}, a stellar mass ($\rm M_*$) of $\log(\rm M_*/M_\odot) = 10.64$, and a SFR of $8.7\ \rm M_\odot\ \rm yr^{-1}$ (W4 band, \textcolor{blue}{van de Sande et al. in prep.}). This results in a specific star formation rate (sSFR) of $0.19\rm\ Gyr^{-1}$, roughly $2-4\times$ above the main sequence. This galaxy is known for its large-scale ionised outflow, first reported by \cite{Lehnert1996} and \cite{Veilleux2003}, who identified a characteristic cross-like structure of gas being expelled from the disc. Recent work by \cite{Elliott2026} has also discussed extraplanar emission in ESO~484-036, reporting the biconical outflow.

Its spectral classification is uncertain due to the high obscuration produced by the dusty disc, with the extinction reaching $\rm E(B-V) \sim 2$ at the galactic centre, making it difficult to determine whether the nuclear activity is powered by an AGN or a starburst. Both \cite{Lehnert1996} and \cite{Veilleux2003} argued that ESO~484-036 must host a starburst-driven outflow, due to morphological and ionisation properties in the extraplanar gas. Galaxies with sSFR greater than $0.1\ \rm Gyr^{-1}$ typically have outflows \citep{Heckman2017,ReichardtChu25}, and with this galaxy being a 2 to 4$\times$ outlier above the star formation rate main sequence (SFMS), it is similar to other galaxies that present SF-driven outflows \citep[e.g.,][]{McPherson2023}. 

However, ESO~484-036 was classified as an obscured (Type 2) AGN by \cite{Zaw2019} using 6dFGS spectra, a result later updated to a mixture of Type 2 AGN and Seyfert II characteristics by \cite{Chen2022}, with both studies using a BPT diagnostic. Although the Million Quasars (Milliquas) catalogue also lists the source as a Seyfert II with potential variability \citep{Flesch2023}, recent analysis of the Zwicky Transient Facility (ZTF) light curves found no evidence of such fluctuations \citep{Sanchez2023}.

Regardless of its central classification, the edge-on orientation of ESO~484-036 makes it an ideal laboratory for studying multiphase outflows. By analysing the wind energetics, we can place constraints on the underlying driver. This near-perfect geometry provides an unobstructed view of the extraplanar emission and the interactions between gas phases.

\subsection{ALMA CO(1$-$0)}
The Atacama Large Millimetre Array (ALMA) observations were carried out during Cycle~10 (project~2023.1.00698.S, PI: A. Bolatto). 
We obtained spectroscopic observations of the CO(1$-$0) transition (rest frequency 115.271~GHz) along with the underlying Band 3 continuum.
This object was observed using both the full ALMA 12m array and the ALMA compact array, to achieve high angular resolutions while ensuring excellent flux recovery.

The datasets were calibrated, combined and imaged using the ALMA pipeline, as provided by the ALMA Regional Centre staff, and the \textsc{Common Astronomy Software Applications} (\textsc{CASA}) package \citep{CASA}. We imaged continuum emission over the full line-free bandwidth, and subtracted this from the data in the $uv$--plane using the \textsc{casa} task \textsc{uvcontsub}. 
The line and continuum data from the combined ALMA datasets were then cleaned and imaged using the \textsc{casa} task \textsc{tclean} and Briggs weighting with a robust parameter of 0.5 (which should provide the best trade-off between sensitivity and angular resolution). The final angular resolution of the data products is 1.1$^{\prime \prime}$.
The resulting three-dimensional (RA, Dec, velocity) datacubes were produced with channel widths of 10\,km s$^{-1}$, and a pixel size of 0.25$^{\prime \prime}$, resulting in a scale of 83 pc per pixel. 

We produced the moment maps for this work using the Python package \texttt{spectral-cube}\footnote{\href{https://github.com/radio-astro-tools/spectral-cube}{https://github.com/radio-astro-tools/spectral-cube}}. Initially, we reprojected the datacube to have a spectral unit of $\rm km\ s^{-1}$ and converted to brightness temperature units (K). The cube was then spatially convolved to the common beam ($1.16^{\prime\prime}\times0.88^{\prime\prime}$) to ensure a consistent angular resolution across all channels. The smooth-mask technique was used to generate a 3D mask for isolating emission from noise. This process involved spatially smoothing a copy of the cube using a Gaussian 2D kernel with a standard deviation of 1 pixel. The root-mean-squared (RMS) noise was then calculated from a set of channels well outside the emission range ($\sim$30 channels preceding the line emission). This RMS noise map was used to create a signal-to-noise (S/N) cube. A binary mask was created by including all pixels where the $\rm S/N\geq5$. This mask was applied to the spatially smoothed cube, and the resulting masked cube was then used to obtain moment maps over the channel range spanning the full extent of the visually confirmed emission line.

\begin{figure}
    \centering
    \includegraphics[width=0.5\textwidth]{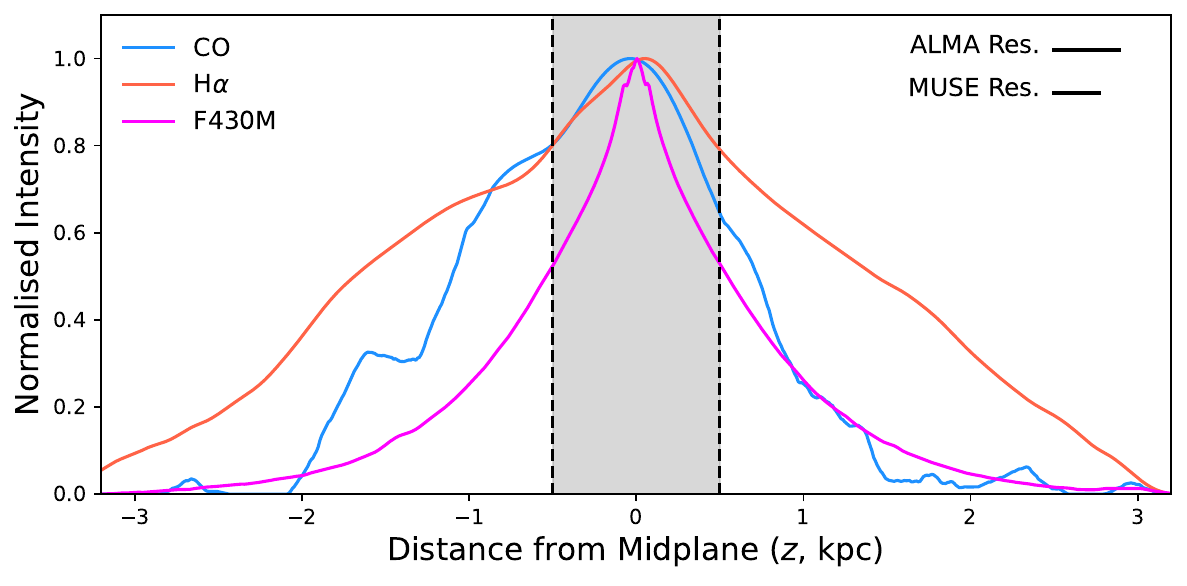}
    \caption{Vertical, normalized intensity profiles for CO (blue), H$\alpha$ (red), and starlight (F430M, magenta). The CO and H$\alpha$ profiles differ clearly from that of the starlight, consistent with extraplanar emission. Grey band is the disc region. Spatial resolutions of ALMA ($1.16^{\prime\prime}\sim 390\rm\ pc$) and MUSE ($0.8^{\prime\prime}\sim 270\rm\ pc$) are showed as black bars for comparison.}
    \label{fig:verticalprofilesDisc}
\end{figure}

\subsection{VLT/MUSE H$\alpha$}
ESO~484-036 was observed as part of the Generalising Edge-on galaxies and their Chemical bimodalities, Kinematics, and Outflows out to Solar environments (GECKOS) project \citep{vandesande2024}\footnote{\url{https://geckos-survey.org/}}. GECKOS is a VLT/MUSE Large Programme (programme~110.24AS, PI: J. van de Sande) targeting 36 edge-on galaxies within a distance range of $10<D<70\,\rm Mpc$ and with a stellar mass within $\pm0.4\,\rm dex$ of the Milky Way \citep[$5\times10^{10}~M_{\odot}$,][]{BlandHawthorn_Gerhard_2016}. The targets are selected from the S$^4$G survey \citep[66\,\%,][]{Sheth2010} and HyperLEDA \citep[34\,\%,][]{Makarov2014}.

Observations with MUSE were performed in wide field mode and using the nominal wavelength range, which provides a field of view of $\rm \sim 1^\prime\times1^\prime$, a spectral sampling of 1.25~\AA~per~pixel over the $4800-9300$~\AA\ wavelength coverage, and a pixel size of $0.2^{\prime\prime}$, resulting in a scale of 67~pc~per~pixel.
ESO~484-036 was observed using one MUSE pointing and three observing blocks (OBs). One GECKOS OB consists of four object (O) exposures with an integration tim
e of 580~s each, and two sky (S) exposures with an integration time of 180~s each, executed in the order OSOOSO.
We use \texttt{pymusepipe}\footnote{\url{https://pypi.org/project/pymusepipe/}} \citep{Emsellem2022} (version 2.28.2) for MUSE data reduction. This package is a wrapper for the MUSE data processing pipeline \citep{Weilbacher2020} and the ESO Recipe Execution Tool \citep[\texttt{esorex};][]{esorex2015}. We follow the same procedures already described in the literature \citep[e.g.,][]{McPherson2023,Hamel2024,Mazzilli2025}.

We model and subtract the stellar continuum using the \texttt{nGIST} pipeline \citep[v5.3.0\footnote{\url{https://geckos-survey.github.io/gist-documentation/}},][]{Fraser2025}, an enhanced extension of the GIST pipeline \citep{Bittner2019} that enables the creation of continuum-subtracted and line-only cubes. We emphasise that we do not use properties derived from the model.
The CONT module of \texttt{nGIST} applies \texttt{pPXF} full spectral fitting \citep{CappellariEmsellem2004,Cappellari2017} to model binned MUSE spectra. Spaxels are binned to $\rm S/N=7$ using Voronoi adaptive binning, with differential stellar population models from \citet{Walcher2009}. Fits are performed over $4750-7100$~\AA\ using 13th-order multiplicative polynomials. Milky Way extinction is included via the \citet{Cardelli1989} law. The best-fit continuum in each Voronoi bin is rescaled to each spaxel and subtracted at the native spatial resolution of MUSE.
The final maps used in this work have a spatial resolution of $\sim0.8^{\prime \prime}$ and are cleaned using $\rm S/N=5$.

\subsection{JWST NIRCam F150W, F200W and F430M}
Infrared imaging observations of ESO~484-036 were obtained with JWST during Cycle 3 (GO project~5637, PI: B. Mazzilli Ciraulo) using the MIRI and NIRCam instruments with the objective to cover the full extent of the disc and detect the faint dust emission from the outflow.

Since our observations cover both the very bright galaxy disc region and the fainter emission in the wind, an observing strategy was employed to mitigate saturation. For NIRCam observations, the FULL sub-array with the MEDIUM2 readout pattern and the SUB640 sub-array with the RAPID readout pattern were used for the wind and the disc, respectively. The full dataset includes data from three MIRI filters: F770W, F1130W, and F2100W, and four NIRCam filters: F150W, F200W, F335M, and F430M. In this work we only make use of F150W, F200W and F430M to create a starlight composite image (Fig.~\ref{fig:RGB}), define disc extension and calculate galaxy inclination. Observations tracing dust in the outflow will be described and discussed in the follow-up work by \citnp{Hernández-Yévenes et al.}.

The uncalibrated NIRCam data for filters F150W, F200W, and F430M were processed using the JWST pipeline version 1.16.1 and the CRDS context \texttt{jwst\_1303.pmap}. In Stage 1 of the pipeline, the default parameters were used with one modification: the \texttt{clean\_flicker\_noise} step was enabled for the FULL sub-array observations to correct for $1/f$ noise, but it was set to False for the SUB640 sub-array observations. Additionally, in the FULL sub-array data for F150W and F200W (specifically the detector B4 files), wisps caused by stray light were visible. These features were removed by appropriately scaling and subtracting wisp templates from the science \texttt{\_cal} files. For Stage 2, the default pipeline parameters were used. However, several artifacts present in the flat fields of the F430M SUB640 sub-array images were manually flagged. In Stage 3, the \texttt{TweakReg} step was skipped for all processed filters because the observed field contains only one to two Gaia stars, which are insufficient for a reliable world coordinate system adjustment. Final images for F150W, F200W, and F430M are background subtracted before making use of them.

\subsection{Spatial definition of the extraplanar and disc regions}

In the bottom panel of Fig.~\ref{fig:RGB}, a clear spatial co-location is observed between the CO and H$\alpha$ emission, both of which extend significantly beyond the underlying stellar continuum. Most of the extraplanar emission is contained within the defined outflow regions (indicated by the blue/red rectangles to distinguish between gas phases), reaching distances well above the disc region (magenta rectangle). In the following, we justify the geometric selection of these regions and the physical criteria used to distinguish the disc from the extraplanar components.

In Fig.~\ref{fig:verticalprofilesDisc} we present vertical profiles of CO(1$-$0), H$\alpha$ and starlight using the F430M observations. The starlight profile tracks the width of the disc. These profiles are derived by integrating the total flux along the minor axis within a region extending $\pm1.5$~kpc from it. This was done to consider the region of interest where the ionised outflow of ESO~484-036 was detected by \cite{Veilleux2003}. The CO emission shows an extended component toward both north and south, with the former being weaker. H$\alpha$ presents a broader profile compared to the disc. Both features are compatible with extraplanar emission that is not associated with the disc, but is instead best explained as an outflow.

Importantly, these extended profiles cannot be attributed to instrumental resolution. The spatial resolution of the MUSE data ($\sim 0.8^{\prime\prime}$, $\sim 270\rm\ pc$) and the ALMA beam ($1.16^{\prime\prime} \times 0.88^{\prime\prime}$, $390 \times 300\rm\ pc$) are significantly narrower than the kiloparsec-scale extent of the observed H$\alpha$ and CO emission. This confirms that the detected features represent genuine extraplanar emission rather than artifacts of the point spread function (PSF) or beam-smearing.

Based on our observations, these vertical profiles, and previous descriptions of the extraplanar emission by \cite{Veilleux2003}, we define a region of interest centred on the galactic nucleus, measuring 3~kpc in width ($\pm1.5$~kpc from the minor axis) and 7~kpc long ($\pm3.5$~kpc from the major axis). This covers the full spatial extent of the observed $\rm CO$ and $\rm H\alpha$ extraplanar emission. To isolate the potential outflow signal, we exclude the central disc region (as defined in Sect.~\ref{sec:discregion}), ensuring that our analysis focuses exclusively on gas residing above the disc midplane.

\section{Multiphase extraplanar emission detected in ESO~484-036}\label{sect:describing_obs}

\begin{figure}
    \centering
    \includegraphics[width=0.48\textwidth]{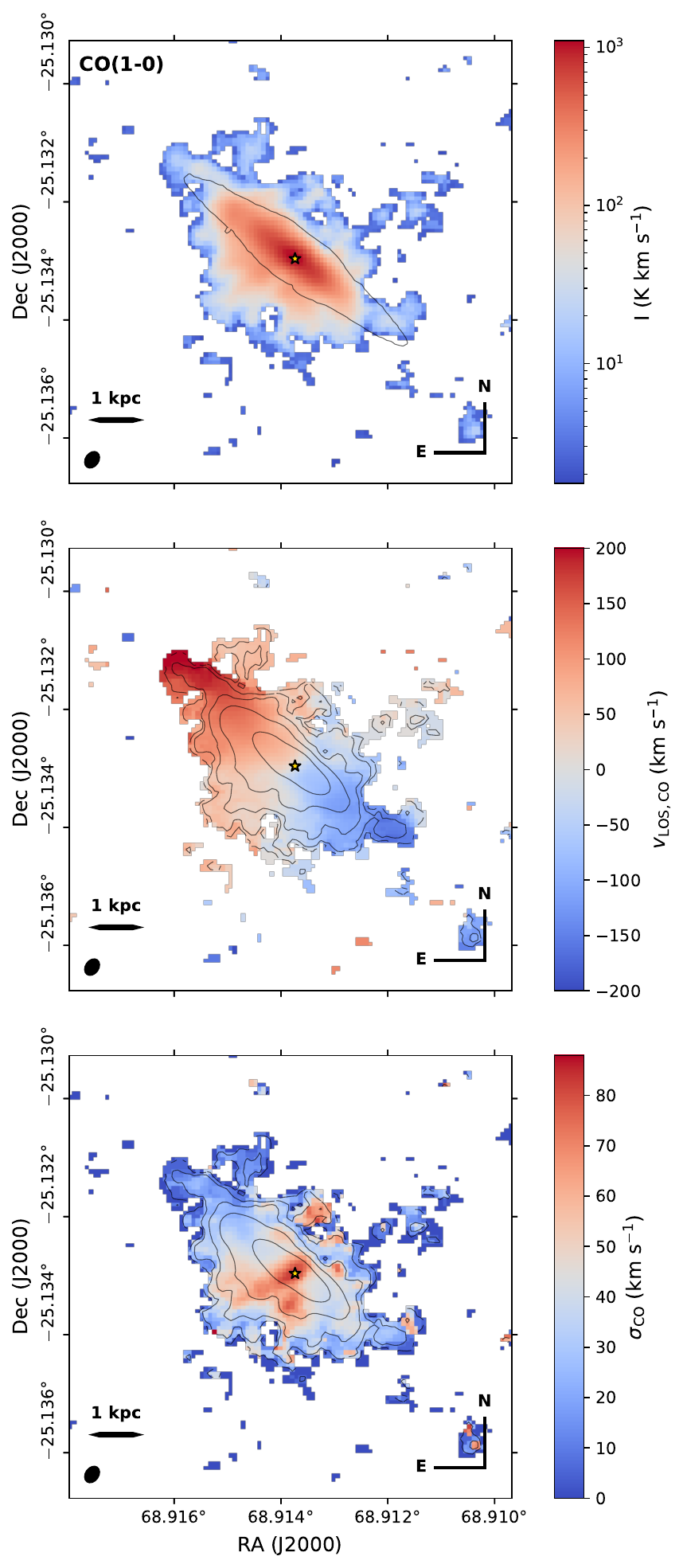}
    \caption{CO(1$-$0) moment maps of ESO~484-036, created from the ALMA data cube with a $\rm S/N=5$ threshold. Top: Intensity (moment 0) shows strong gas emission in the south, with a small decrement near the minor axis; the north lacks clear structure. A starlight (F200W) contour is overlaid. Middle: Line-of-sight velocity (\vlos, moment 1) indicates a wind preserving some disc rotation. Velocities are shown in the systemic rest frame of the galaxy. Bottom: Velocity dispersion ($\sigma$, moment 2) shows a southern V-structure, consistent with outflow kinematics. CO contours are overlaid in both centre and bottom panels.}
    \label{fig:CO_moments}
\end{figure}

\begin{figure}
    \centering
    \includegraphics[width=0.48\textwidth]{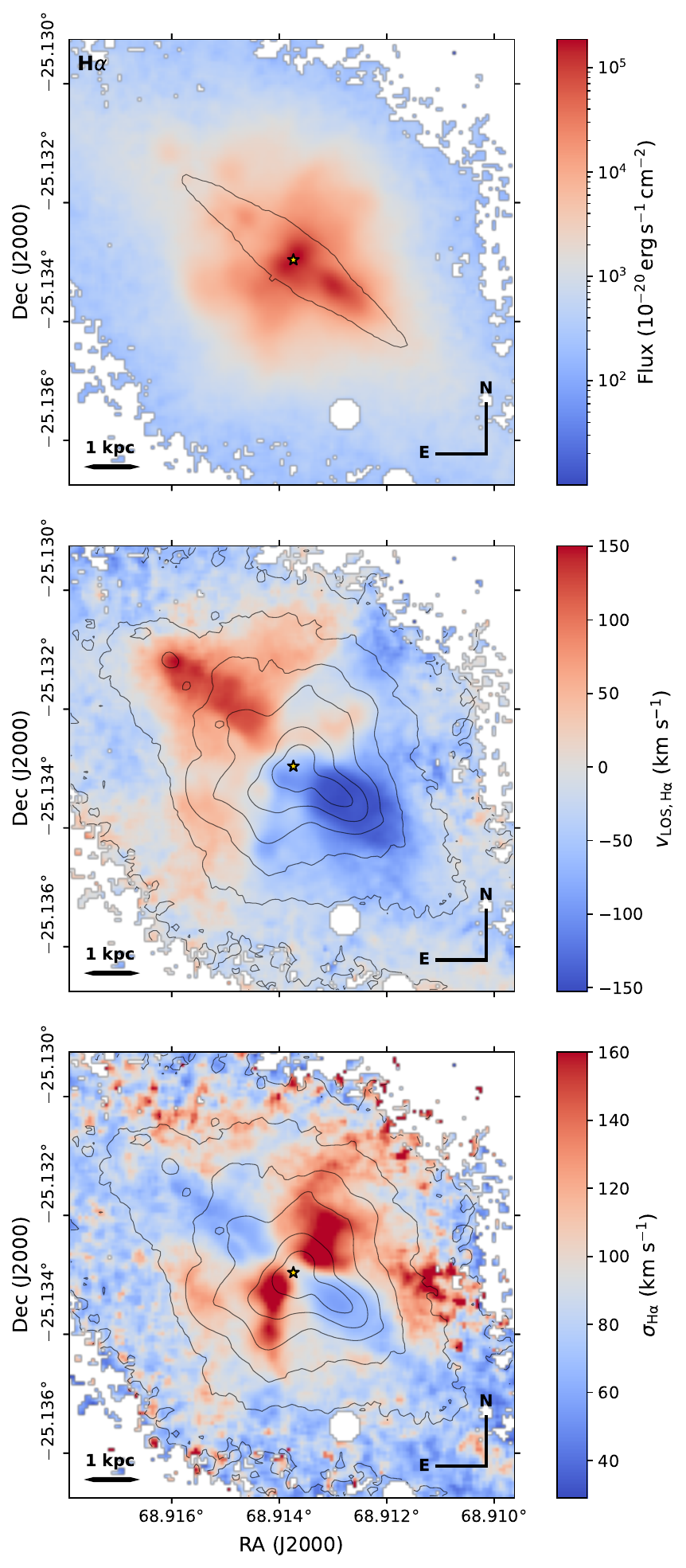}
    \caption{H$\alpha$ flux and kinematics maps of ESO~484-036 created from the VLT/MUSE data using a $\rm S/N=5$ threshold. Top: Flux map displays a prominent biconical extraplanar emission extending more than $\sim 3$~kpc to each side. A starlight (F200W) contour is overlaid. Middle: Line-of-sight velocity ($\upsilon_{\rm LOS}$) map indicates that the extraplanar emission preserves a component of the underlying disc rotation. Offsets from the midplane and expected zero-velocity in the centre can be attributed to the high obscuration. Velocities are shown in the systemic rest frame of the galaxy. Bottom: Velocity dispersion ($\sigma$) map corrected for instrumental broadening. The colorbar is capped at $160\rm\ km\ s^{-1}$ to highlight the high-dispersion biconical structure, consistent with outflow kinematics and similar to the top panel. $\rm H\alpha$ flux contours are overlaid in both centre and bottom panels.}
    \label{fig:Halpha_maps}
\end{figure}

\subsection{Molecular gas detected with CO(1$-$0)}\label{sec:molecular_outflow_maps}
The CO flux extends away from the galaxy on the minor axis beyond the stellar disc, as shown in the top panel of Fig.~\ref{fig:CO_moments}. We observe strong emission on the southern side of the disc, where two broad molecular gas structures extend from the galaxy centre to $\gtrsim2$~kpc below the midplane. A central region aligned with the minor axis separates these structures and shows an intensity approximately half that of extended emission. On the northern side, we clearly detect extended CO emission up to $\sim2.5$~kpc, although much weaker than the southern emission. We find very little substructure within the northern extended component. Resolution limits further interpretation.

Line-of-sight velocities (\vlos) show stable rotation of the molecular gas without strong perturbations, as shown in the middle panel of Fig.~\ref{fig:CO_moments}. The \vlos\ ranges from +200 to -200~$\rm km\ s^{-1}$. The extraplanar gas also shows signatures of rotation, albeit at lower velocity, suggesting that the gas in the region keeps some rotation of the disc.
We use the velocity offset between this extraplanar gas and the projected disc rotation to derive the intrinsic vertical outflow velocity profiles and perform the geometric deprojections detailed in Sect.~\ref{sec:dynamics}.

Emission north of the midplane is slightly blueshifted, relative to the systemic velocity of the galaxy, while emission south of the midplane is redshifted. If this extended emission traces an outflow, this is explained by the gas simply moving toward or away from the observer, depending on the galaxy orientation.

High velocity dispersions ($\sigma_{\rm CO}$) are observed in the extraplanar emission, with a clear V-shape structure at the south of the midplane, as shown in the bottom panel of Fig.~\ref{fig:CO_moments}. The contours indicate that this is spatially correlated with the brightest CO regions within the southern extraplanar emission. High $\sigma_{\rm CO}$ measured in the southern region likely arises from a mix of unresolved velocity gradients, overlapping kinematic components along the line-of-sight, and projection effects associated with the edge-on geometry. The characteristic V-shape is highly consistent with the projected geometry of a biconical outflow. In this edge-on orientation, such a structure naturally produces an enhanced $\sigma_{\rm CO}$ as the line-of-sight intersects multiple expanding shells of the cone, particularly near the outflow boundaries \citep[e.g.,][]{Salak2020, Cortese2026}. 

\subsection{Ionised gas detected with H$\alpha$}\label{sec:ionised_outflow_maps}

In the top panel of Fig.~\ref{fig:Halpha_maps} we present the H$\alpha$ flux map of ESO~484-036. We resolve the extraplanar structure originally reported by \cite{Veilleux2003}, now with better spatial resolution. A symmetric X-shaped structure extends up to 3~kpc on both sides of the galactic midplane. A bright emission-line region lies in the galactic midplane west of the minor axis, with peak fluxes comparable to those at the centre. A heavily obscured galactic nucleus could cause this effect, creating the apparent peak on the south-western side, near the hidden nucleus. Another possible explanation is a SF region, but the lack of extraplanar emission coming from it makes this scenario less probable.

The middle panel of Fig.~\ref{fig:Halpha_maps} presents \vlos\ of the ionised gas. In the galaxy disc, high obscuration ($A_V \sim 4.5-6$ mag) prevents a direct measurement of the true internal gas rotation. Since the dense dust column hides the high-velocity gas at the midplane, the H$\alpha$ emission we detect originates from the less-attenuated outer layers of the disc. This results in a peak rotation that is offset from the true stellar midplane defined by the near-IR F200W band. We provide an analysis of these extinction effects in Sect.~\ref{sec:ionisedmass}. See also \cite{Fraser2025} and \cite{Rutherford2025} for a discussion of extinction effects on kinematics of edge-on galaxies.

The impact of extinction on disc velocity will also affect measurements of outflow properties. Since we isolate the outflow kinematics by subtracting a disc rotation curve from the extraplanar gas, any dust-induced skew in the perceived midplane velocity directly propagates into the outflow calculations. For the ionised phase, midplane obscuration introduces a systematic shift of approximately $30\rm\ km\ s^{-1}$ in the final velocity profiles (Sect.~\ref{sec:dynamics}).

As observed in the molecular gas, extraplanar emission in the north side presents a slight blueshift, explained by the galaxy being inclined toward the observer on the northern side.

The bottom panel of Fig.~\ref{fig:Halpha_maps} presents the velocity dispersion map of the ionised gas ($\sigma_{\rm H\alpha}$). A clear biconical, extraplanar structure correlates high $\sigma_{\rm H\alpha}$ regions with the observed extraplanar emission. These kinematics closely mirror those of the molecular gas shown in Fig.~\ref{fig:CO_moments}. The same physical interpretation applies, the high dispersion arises from a mix of unresolved, complex kinematics as the ionised gas is expelled from the galactic midplane.

\subsection{Spatial correlations between CO and H$\alpha$ morphology and kinematics}

The biconical ionised gas originates from the same region at the galactic centre as the molecular gas, showing a spatial correlation between the ionised and molecular phases in both morphology and kinematics, especially on the south side of the midplane (compare Figs.~\ref{fig:CO_moments} and~\ref{fig:Halpha_maps}). 

\begin{figure}
    \centering
    \includegraphics[width=0.5\textwidth]{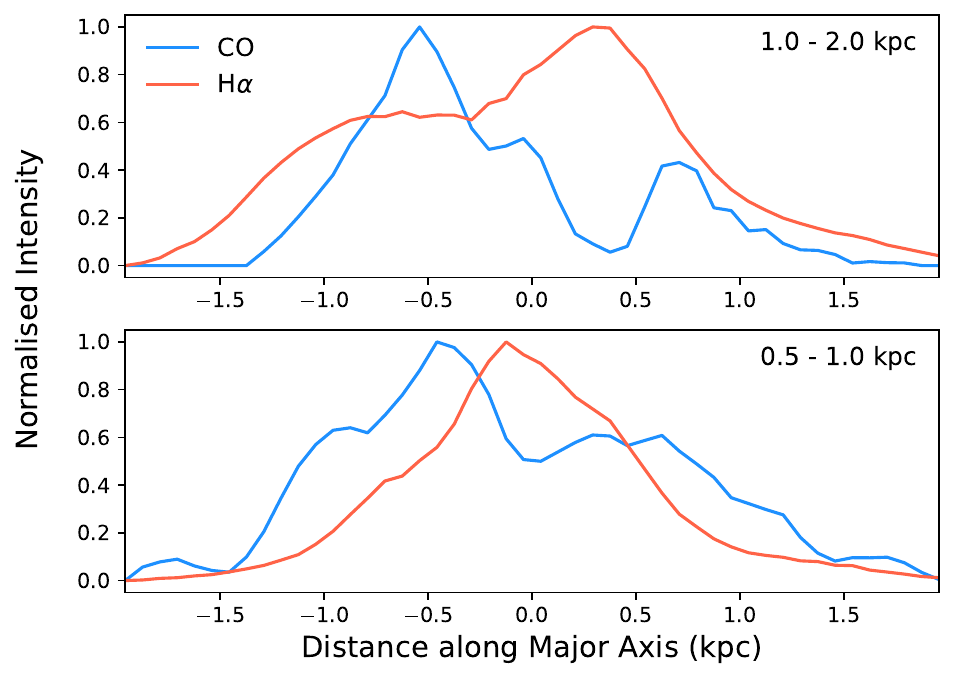}
    \caption{Horizontal, normalised profiles for CO (blue) and H$\alpha$ (red) maps for the south outflow at different heights; $z=0.5-1\ \rm kpc$ (bottom) and $z=1-2\ \rm kpc$ (top). Near the disc midplane ($z < 1$~kpc), the ionised gas peaks centrally where the molecular emission shows a local minimum, consistent with an ionised core surrounded by a molecular sheath. At larger heights ($z > 1$~kpc), both profiles broaden, reflecting the biconical outflow expansion. Profiles are normalised to their local maxima to emphasise the spatial anti-correlation between the ionised and molecular phases.}
    \label{fig:HorizontalCuts}
\end{figure}

The observed behaviour of the ionised and molecular gas in ESO~484-036 is consistent with the description of biconical, starburst-driven outflows in the literature \citep{Leroy2015, Nguyen2022}, which features a central ionised phase originating from an extended base, surrounded by molecular gas. Figure~\ref{fig:HorizontalCuts} illustrates this multiphase distribution. We show two horizontal profiles for the south outflow of ESO~484-036, showing the median flux at different distances from the galaxy midplane. Each flux profile is normalised, so the maximum in the plotted region is 1 and the minimum is 0, allowing direct comparison of local minima and maxima between phases.

In the gas at $z=0.5-1\ \rm kpc$, the H$\alpha$ flux peaks centrally, while the CO flux shows a double-peaked profile, with the H$\alpha$ spatially correlating to a local minimum in CO. If this emission is indeed an outflow, then this suggests that the molecular phase surrounds the ionised phase at the outflow base. This is the expected behaviour of a multiphase wind \citep{Veilleux2020}. In the gas at $z=1-2\ \rm kpc$, the H$\alpha$ profile broadens, presenting a peak that correlates with a CO minimum at $\sim0.4\ \rm kpc$ and presenting a shelf on the left side of the profile. This is consistent with a conical outflow, opening as the gas travels out of the disc. 

The spatial distribution of the ionised and molecular phases suggests a direct dynamic link between the gas reservoirs. As shown by the contours in Fig.~\ref{fig:CO_moments}, the peaks in $\sigma_{\rm CO}$ correlate with the regions of highest CO intensity in the southern outflow. Ionised gas shows a similar correlation in the bottom panel of Fig.~\ref{fig:Halpha_maps}. Note that the kinematic structures described by the high $\sigma$ in both phases are anti-correlated, with peak $\sigma_{\rm H\alpha}$ being centrally located, and $\sigma_{\rm CO}$ presenting a V-shape, with minimum dispersion near the minor axis.

This suggests a scenario in which the hot, ionised wind entrains the molecular component as it expands. The high $\sigma_{\rm CO}$ represents the kinematic signature of this entrainment process, where the ionised phase transfers momentum and turbulence to the denser cold phase.

All features observed in the molecular and ionised gas can be explained by an outflow in the shape of a bi-frustum, i.e., a cone with its top cut off by a plane parallel to the base. This outflow is ejected from an extended central region in the nucleus where the starburst resides, according to the central starburst model described in \cite{Leroy2015}. This can also be a nuclear ring created by a bar that deposits cold gas into central regions \citep{Nguyen2022,Fraser2025}. Most of the gas is ejected roughly perpendicular to the line-of-sight (see our comment about tilt of outflow above), but the kinematic signatures are consistent with an outflow that retains some rotation above the midplane. 

\section{Constraining the extraplanar emission geometry}\label{sec:geometry}

If we attribute the extraplanar emission to an outflow, calculating its properties requires correctly determining three angles: position angle (PA), inclination ($i$), and half-opening angle ($\theta$). An accurate PA is essential for defining the major and minor axes, which in turn define the outflow base and allow us to calculate the velocity offsets from the disc rotation. Both $i$ and $\theta$ are necessary to describe the geometry of the outflow. This is used to deproject the velocity offsets, a crucial step to estimate the vertical velocity and the properties of the outflow.

\subsection{Galaxy position angle and inclination}

The GECKOS survey provides a position angle of $\rm PA \sim 52^\circ$ for ESO~484-036 (\textcolor{blue}{van de Sande et al. in prep.}). A systemic velocity of $\upsilon_{\rm sys} = 5058\rm\ km\ s^{-1}$ is available in the literature \citep{Makarov2014}. We cross-checked these global parameters using the PaFit package \citep{Krajnovic2006} in our CO velocity field. This returned $\rm PA = 52.0 \pm 3.5^\circ$ and $\upsilon_{\rm sys} = 5055.6 \pm 3.6\rm\ km\ s^{-1}$, providing consistent measurements.
To independently verify the kinematic centre, PA and $\upsilon_{\rm sys}$, we used intensity map (moment 0) and \vlos\ (moment 1) from the CO data. 

Using this value of PA, we tested the rotation symmetry. This method assumes that at equal radial distances ($r$) along the main axis, the velocities of the receding ($\upsilon_{+r}$) and approaching ($\upsilon_{-r}$) sides are symmetric: $\upsilon_{+r} + \upsilon_{-r} - 2 \upsilon_{\rm sys} = 0$. We define a quantitative asymmetry metric by extracting velocity profiles along the adopted PA at equal positive and negative radii from a test centre. The metric is $\left| (\upsilon_{+r} - \upsilon_{\rm sys}) + (\upsilon_{-r} - \upsilon_{\rm sys}) \right|$, i.e. the deviation from an ideal symmetric profile. We applied a flux mask ($I \ge 0.3 I_{\rm max}$) to sample only high-flux regions near the centre of the molecular disc to exclude extraplanar emission. 
We performed a grid search to minimise this asymmetry metric, refining the central coordinates and $\upsilon_{\rm sys}$ through an iterative search. For the coordinates $({i}_{x}, {i}_{y})$, we calculated the asymmetry metric, leading to a $\upsilon_{\rm sys}$ that minimises this metric for those specific coordinates.

This minimisation procedure yielded a best-fit kinematic centre $(i_x, i_y)$ (indicated by a gold star in Figs. $\ref{fig:RGB}$, $\ref{fig:CO_moments}$, and $\ref{fig:Halpha_maps}$) and a refined systemic velocity of $\upsilon_{\rm sys} = 5053.5\rm\ km\ s^{-1}$. Although this value is consistent with the literature ($5058\rm\ km\ s^{-1}$), we adopt our derived $\upsilon_{\rm sys}$ as it provides the most internally consistent kinematic result from the resolved observations.

After identifying the kinematic centre, we defined the major and minor axes by finding the single PA that best satisfies both photometric and kinematic constraints. We aligned the major axis with the high-intensity disc emission while simultaneously fitting the minor axis to the vertical region where $|\upsilon_{\rm LOS} - \upsilon_{\rm sys}|\approx 0 \rm\ km\ s^{-1}$. This approach yields a $\rm PA = 54.8^\circ$, which closely matches the H$\alpha$ flux morphology and the stellar disc orientation seen in the JWST starlight composite.

Our derived position angle of $\rm PA = 54.8^\circ$ is in agreement with the global GECKOS value and shows a clear alignment with both the H$\alpha$ flux morphology and the JWST stellar disc. We adopt $\rm PA = 54.8^\circ$ for all subsequent analyses given the consistency across our observations.

The GECKOS survey provides an $i \sim 87.5^\circ$ for ESO~484-036, coming from the average of multiple approaches by multiple members of the GECKOS team. To independently verify this, we fit an ellipse to the brightest region of the disc in the JWST starlight images (F150W and F200W), selected using the $75^{\rm th}$ percentile as a threshold via \texttt{scikit-image measure}. We calculated $i \sim 84^\circ$ using Equations 1 and 2 from \citet[assuming $T=2$]{Bottinelli1983}. Varying the threshold ($70^{\rm th}$ to $99^{\rm th}$) yielded $82^\circ \le i \le 88^\circ$. 

We adopt $i = 87.5^\circ$ as our primary inclination for the kinematic model. However, given that deprojected outflow properties are highly sensitive to the exact orientation in edge-on systems, this motivates a systematic sensitivity analysis. In Appendix~\ref{app:variations}, we quantify how the lower inclination ($i = 84^\circ$) would affect the final outflow properties, ensuring that our physical conclusions remain robust across a range of plausible geometries.

\begin{figure}
    \centering
    \includegraphics[width=0.5\textwidth]{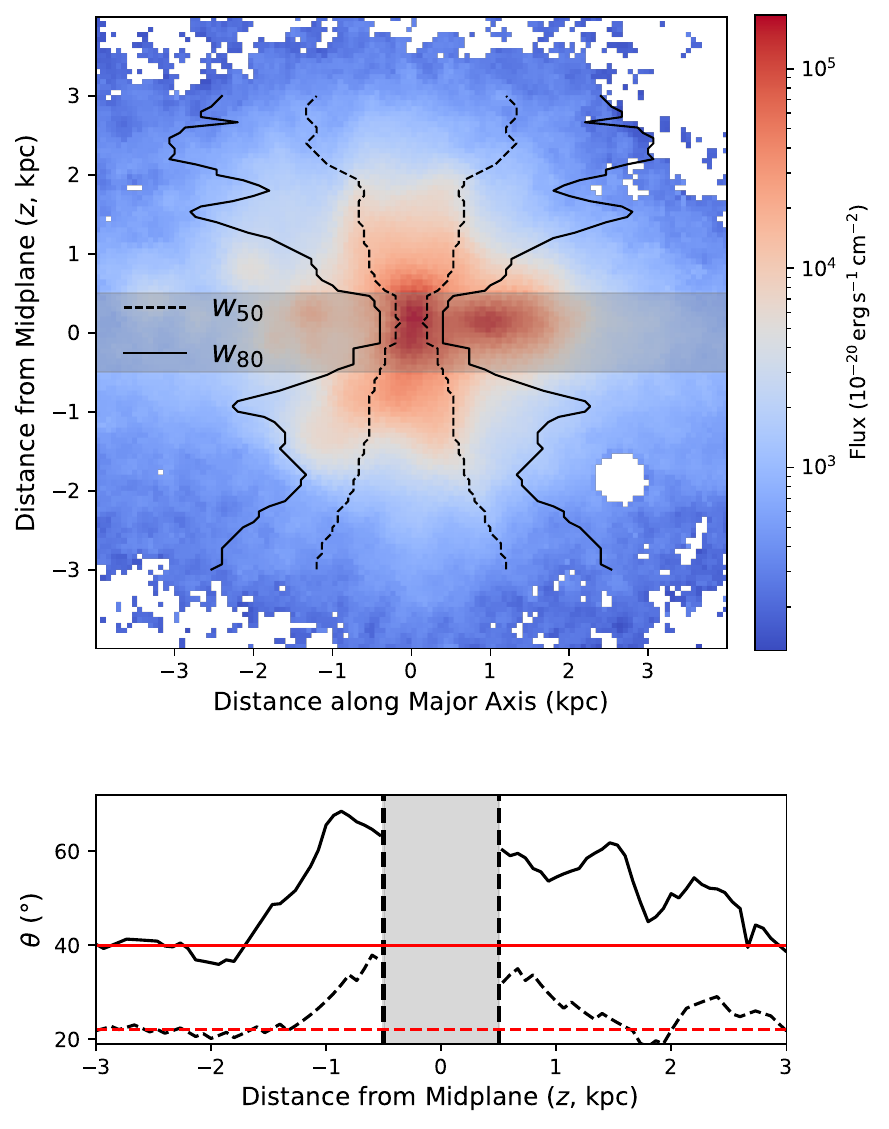}
    \caption{Top: Outflow geometry determined using the method from \protect\cite{McPherson2023}, overlaid on a rotated H$\alpha$ flux map. Using the minor-axis as a centre, the outflow region is determined by $50\%$ ($w_{50}$, black dashed line) and $80\%$ ($w_{80}$, black solid line) widths of the flux. Bottom: Half-aperture angle ($\theta$) calculated for this geometry against distance from midplane. $\theta$ is in the range of $22^\circ-40^\circ$, represented by the red dashed and solid lines of convergence for each contour, respectively. Grey band in both panels is the disc region.}
    \label{fig:Geometry}
\end{figure}

\begin{figure*}
    \centering
    \includegraphics[width=\textwidth]{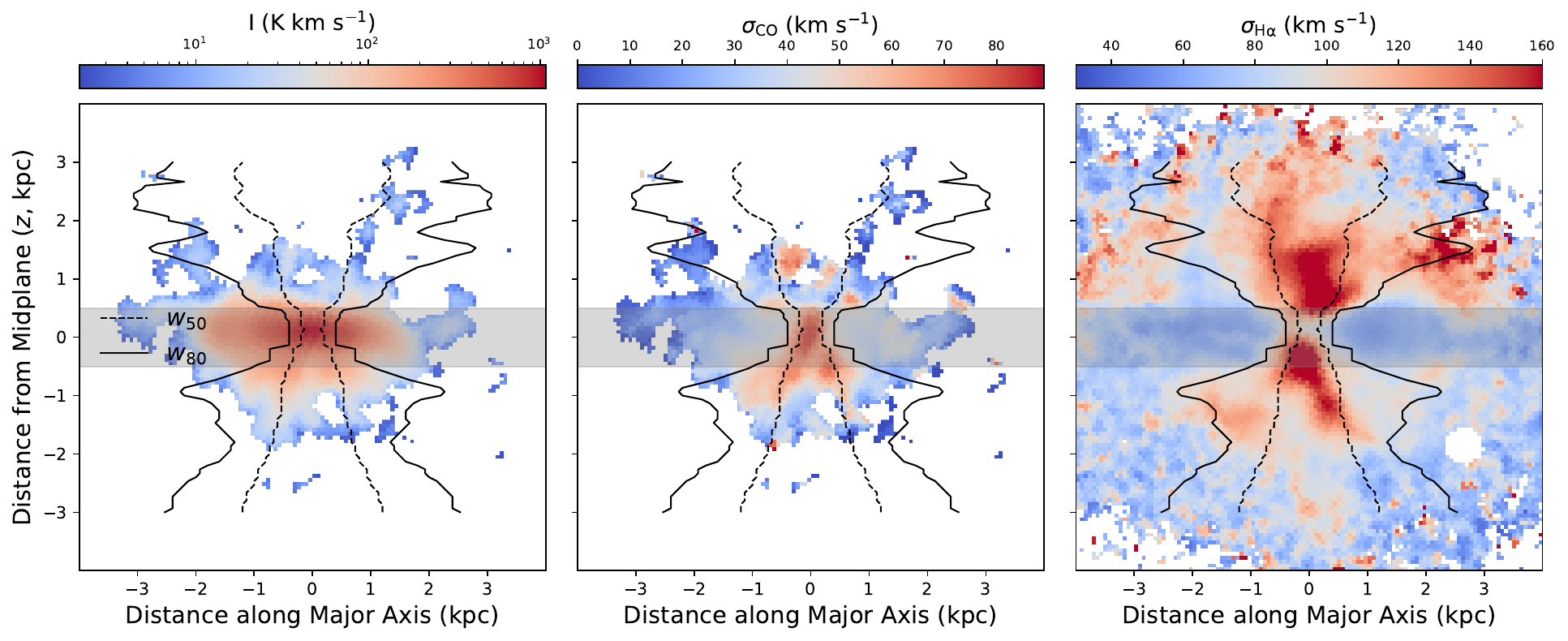}
    \caption{Outflow geometry derived from the H$\alpha$ flux map (Fig.~\ref{fig:Geometry}) plotted on top of CO intensity (moment 0, left), CO velocity dispersion ($\sigma_{\rm CO}$, middle) and H$\alpha$ velocity dispersion ($\sigma_{\rm H\alpha}$, right). Molecular gas clouds in the south outflow mostly reside in between the $w_{50}$ and $w_{80}$ widths, with the local minimum of CO (see Fig.~\ref{fig:HorizontalCuts}, bottom panel) residing in the inner cone. Peak $\sigma_{\rm CO}$ align with the $w_{50}$ width. Peak $\sigma_{\rm H\alpha}$ describes a kinematic structure that is completely enclosed by the $w_{50}$ width.}
    \label{fig:Geometry_top_others}
\end{figure*}

\subsection{Outflow base and disc extension}\label{sec:discregion}

To distinguish the outflow base from the stellar component, we analysed the vertical light distribution using JWST imaging (F150W, F200W, and F430M filters). We extracted vertical luminosity profiles perpendicular to the major axis and normalised them by the total flux within the central 3~kpc ($\pm1.5$~kpc from the minor axis). We then calculated the spatial extent of the disc by symmetrically expanding from the galactic midplane until the enclosed area contained 50\% of the total integrated flux. Although the widths vary across wavelengths (0.57 to 0.44 kpc), this method consistently yields compact vertical widths. To ensure the complete exclusion of stellar contamination from our outflow kinematics, we defined the disc region as a rectangle 3~kpc wide ($\pm1.5$~kpc from the minor axis) by 1~kpc long ($\pm0.5$~kpc from the major axis). This region is shown in the top right panel of Fig.~\ref{fig:RGB} as a magenta rectangle and as a grey shaded region in all the vertical profiles presented in this work. One of the profiles used in this analysis (F430M) is shown in Fig.~\ref{fig:verticalprofilesDisc}.

\subsection{Outflow half-opening angle}

We calculate the half-opening angle ($\theta$) following the approach of \cite{McPherson2023}, applying it to our H$\alpha$ flux map to constrain $\theta$. We present the geometry in Fig.~\ref{fig:Geometry}. Assuming outflow symmetry, we fixed the outflow axis to our previously calculated minor axis. We calculated the $50\%$ and $80\%$ spatial width contours ($w_{50}$ and $w_{80}$) along the minor axis, which enclose the respective cumulative percentage of the H$\alpha$ flux.
We used $w_{80}$ because higher percentage contours showed unstable behaviour in this dataset. These calculated regions appear in the top panel of Fig.~\ref{fig:Geometry}, from which we derive $\theta$ using simple trigonometry. In the bottom panel of Fig.~\ref{fig:Geometry} we present how $\theta$ changes with distance from the midplane for our calculated contours. For $w_{50}$, the contour clearly forms a biconical shape with a stable aperture. Both the south and north converge to approximately $\theta_{\rm min}\sim22^\circ$, which we adopt as the narrowest possible aperture for the outflow. In the case of $w_{80}$, the contour behaves more erratically and only converges to a stable $\theta$ in the most distant southern region between $2-3$~kpc, where it converges to $\theta_{\rm max}\sim40^\circ$. We note that while this represents the most distant and lowest S/N region of our detection, this convergence may reflect a genuine structural broadening of the outflow. We adopt this as the broadest possible aperture for our outflow.

Figure~\ref{fig:Geometry_top_others} presents the $w_{50}$ and $w_{80}$ contours overlaid on the CO intensity, CO velocity dispersion ($\sigma_{\rm CO}$), and H$\alpha$ velocity dispersion ($\sigma_{\rm H\alpha}$) maps, respectively. In the left panel, the two molecular clouds surrounding the inner ionised gas reside between these two contours in the southern outflow. The middle panel shows the V-shaped kinematic structure of the southern molecular gas aligned with the $w_{50}$ contour. In the right panel, the $w_{50}$ contour encloses the biconical ionised kinematic structure. All of these further justify our choice of this geometry for the outflow model.

Although the $w_{50}$ contour is fully enclosed in the outflow region, the $w_{80}$ contour reaches a maximum width of $3$~kpc (from the minor axis), twice the width of our rectangular outflow region defined in Fig.~\ref{fig:RGB}. The $w_{80}$ contour contains $4\%$ more flux than the rectangular outflow region. This difference is insufficient to motivate a change in region definition, as the simpler rectangular region already captures the bulk of the extraplanar emission.

\begin{figure}
    \centering
    \includegraphics[width=0.5\textwidth]{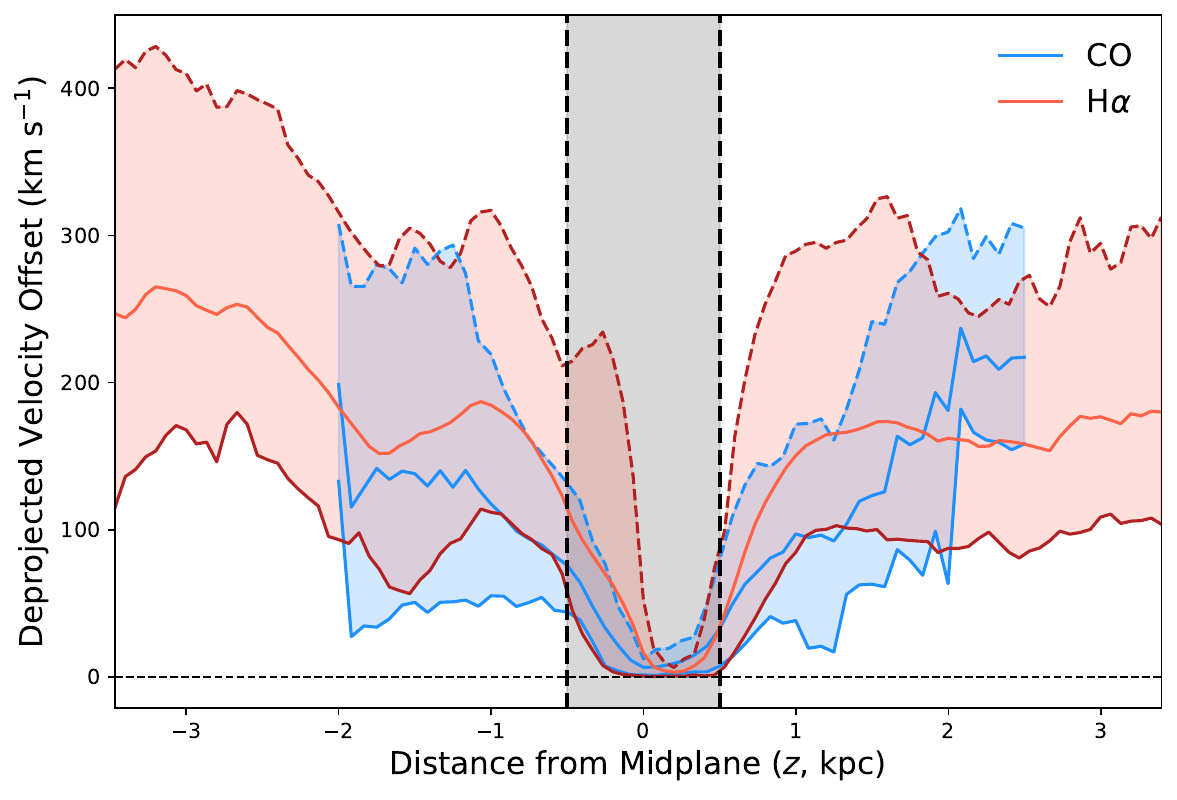}
    \caption{Velocity profiles for the molecular (blue) and ionised (red) phases of the outflow in ESO~484-036. These are calculated by subtracting the rotation of the disc to the \vlos\ maps, excluding the centre region near the rotation axis, taking the median of the velocity offsets and then deprojecting it. Median velocity profiles are deprojected using $\theta=30^\circ$. To show the complete range of probable velocities, 16$^{\rm th}$ and 84$^{\rm th}$ percentile curves (solid and dashed grey curves, respectively) of the velocity offset profiles are deprojected by $\theta_{\rm max}=40^\circ$ and $\theta_{\rm min}=22^\circ$, respectively. Grey band is the disc region. Given the better coverage of the ionised phase, the profiles extent up to 3 kpc in both sides of the disc. south present similar velocities compared to the cold phase at 1.5 kpc, but after 2 kpc, it goes at least 100 $\rm km\ s^{-1}$ higher (median profile).}
    \label{fig:velprofs}
\end{figure}

\section{Kinematics of the extraplanar emission}\label{sec:dynamics}

\subsection{Velocity profiles of the extraplanar gas}\label{sec:velprofs_nolag}

We present the final velocity profiles in Fig.~\ref{fig:velprofs}. We provide a range of profiles using $\theta_{\rm min}$ and $\theta_{\rm max}$ from Sect.~\ref{sec:geometry}. Dashed lines represent the highest possible velocity profile under our assumed geometry, highlighting high uncertainty from deprojection.

Due to the high inclination of ESO~484-036 ($i=87.5^\circ$), most of the extraplanar gas is expelled nearly perpendicular to the line-of-sight. This geometry introduces significant uncertainties in deriving the true outflow velocity, which propagate non-linearly into the mass, momentum, and energy flux estimates ($\dot{M} \propto \upsilon, \dot{p} \propto \upsilon^2, \dot{E} \propto \upsilon^3$). To isolate intrinsic outflow kinematics, we derived velocity offset maps ($\upsilon_{\rm offset}$) for the molecular and ionised phases by subtracting a disc rotation curve from the observed fields \vlos\ shown in Figs.~\ref{fig:CO_moments} and~\ref{fig:Halpha_maps} and then we deproject those offsets based on the geometry derived in Sect. \ref{sec:geometry}. 

To derive the disc rotation curve, we analysed vertical slices of the \vlos\ map within the disc region ($\pm 0.5$~kpc of the midplane). For each radial bin, we initially identified the pixel with the maximum \vlos. In the nuclear region where $\upsilon_{\rm LOS} \approx 0 \rm\ km\ s^{-1}$ and the peak-velocity criterion is less reliable, we adopted the velocity measured directly at the midplane. Once this primary kinematic curve is established, we calculate the final rotation velocity for each radial bin by averaging the \vlos\ values within a $\pm 0.2$~kpc aperture centred on these identified peak-velocity coordinates. 

This approach effectively mitigates the impact of midplane extinction, which shifts the apparent kinematic centre in the ionised phase. The resulting $\upsilon_{\rm offset}$ is calculated as $|\upsilon_{\rm LOS}(x,z) - \upsilon_{\rm rot}(x)|$. We note that this effectively attributes all of the projected velocity difference to outward motion, which may be an overestimate \citep[see][for a similar case]{Mazzilli2025}. To avoid biasing our results toward low velocities, we mask the region along the minor axis, where the projected rotational and outflow components are intrinsically near zero. These final maps are then used to get profiles of $\upsilon_{\rm offset}(z)$ for each gas phase by taking the median and $\pm1\sigma$ errors.

We define the deprojection factor, $\alpha$, as the proportionality factor that relates the $\upsilon_{\rm offset}$ calculated previously to the true vertical outflow velocity ($\upsilon_{\rm out}$), such that $\upsilon_{\rm out} = \upsilon_{\rm offset}/\alpha$. To calculate it, we need the three angles that describe the outflow geometry; inclination ($i$), half-opening ($\theta$) and azimuth ($\phi$) using the following expression:
\begin{equation}
    \alpha = |\sin\theta\sin i \sin\phi \pm \cos\theta\cos i|
\end{equation}
In our case, we only make use of $i$ and $\theta$ and assume the maximum projection in the direction of $\phi=90^\circ$, given the resolution of our data. The expression simplifies to $\alpha = |\cos(i\pm\theta)|$, depending on the side of the outflow that is being deprojected. We emphasise that deprojection using only $i$, in this specific case, is not reliable, due to the rapid decrease of $\alpha$ for small variations of $i$ close to 90$^\circ$.

To test maximum outflow velocities, we deprojected the peak $\upsilon_{\rm offset}$ ($\sim 160\rm\ km\ s^{-1}$ for CO) into intrinsic outflow velocities ($\upsilon_{\rm out}$) using the biconical geometry described in Sect.~\ref{sec:geometry}. Note that the peak $\upsilon_{\rm offset}$ appears in only a few pixels of the real image, and our profiles never reach these high velocities. For a median opening angle of $\theta=30^\circ$, the deprojected velocities range from $300-350\rm\ km\ s^{-1}$. However, due to the edge-on orientation, the southern outflow, which is inclined away from the observer, is highly sensitive to small geometric variations. Reducing the opening angle to $\theta=22^\circ$ causes the southern deprojected velocity to reach $\sim 480\rm\ km\ s^{-1}$, approaching the escape velocity for a Milky-way-like mass galaxy ($\sim 550\rm\ km\ s^{-1}$; \citealt{Koppelman2021}). We note that ESO~484-036 is slightly more massive than the Milky-way, so a greater $\upsilon_{\rm esc}$ is expected.

We now focus on median velocity profiles, shown as solid lines in Fig.~\ref{fig:velprofs}.

On the north side of the galaxy, CO velocity systematically increases to a maximum of $\sim250$ $\rm km\ s^{-1}$ at 2~kpc. On the southern side, the velocity is relatively constant at $\sim130$ $\rm km\ s^{-1}$, between $1-2$~kpc. This confirmed kinematic asymmetry, together with the observed structural asymmetry, is a common feature in galactic outflows \citep{Krieger2019, McPherson2023, Fisher2025}. 

The continued positive gradient of the molecular gas in the northern emission suggests its ongoing expansion. Although a velocity of $250$ $\rm km\ s^{-1}$ is significant for cold gas, it remains within the typical range of molecular outflow speeds \citep{Veilleux2020}. This contrasts with systems such as Arp~220 or NGC~~4945, where molecular gas velocities, driven by an AGN, can exceed $600$ $\rm km\ s^{-1}$ or even $800$ $\rm km\ s^{-1}$ \citep{Barcos2018, Bolatto2021}. The sustained positive gradient suggests that the injection of kinetic energy is more efficient on the northern side.

Southern CO emission presents an almost constant velocity of $130$ $\rm km\ s^{-1}$ from $1-2$~kpc. It is significantly below the escape velocity, further suggesting that part of the mass expelled may return to the disc. In this scenario, the ejected material stalls and falls back to the galactic disc because its velocity is far below the escape velocity of ESO~484-036, causing the mass flux through successive radial surfaces to decline. This behaviour is reminiscent of the cold gas flow in M82, which remains confined to within a few kiloparsecs and fails to reach the halo \citep{Leroy2015}.

Ionised gas at the northern side exhibits an almost perfectly flat velocity profile between $1-3.5$~kpc with a velocity of $\sim180$ $\rm km\ s^{-1}$. At 1~kpc, it moves $50\%$ faster than the co-located molecular gas. At 2~kpc, molecular gas moves $65\%$ faster than the ionised gas, being this the location where molecular gas reaches its peak velocity and where ionised gas shows a flat profile. At this distance, molecular gas is only barely detected, so the velocity is likely poorly constrained. Possible explanations include that the observed ionised gas is not the main driver (with hot X-ray gas playing that role), or that an earlier outburst accelerated the molecular gas to these velocities.

At 2~kpc in the southern side, the ionised gas presents a velocity of $\sim 200$ $\rm km\ s^{-1}$ that rises to almost $\sim 280$ $\rm km\ s^{-1}$ at 3~kpc. This increase in velocity with distance from the disc is a known feature of ionised gas in galactic superwinds \citep{Thompson2024, Cronin2025}. A simple explanation for this positive spatial gradient is that the gas that was expelled first in the outflow always moves faster than successive layers of gas. This can also suggest that the ionised component continues to be driven or entrained by an underlying energy source \citep{Schneider2020}.

Between $1-2$~kpc in the south of the midplane, the convergence of molecular and ionised gas velocities (to within a few $\rm km\ s^{-1}$) could suggest hydrodynamic interaction. The warmer ionised gas, tracing the fast wind, is decelerated by entrainment with massive cold clouds, consistent with turbulent radiative mixing layers \citep{Fielding2022}. This interaction occurs where we no longer detect molecular gas. Reduced entrainment then allows the ionised gas, still coupled to energy injection, to go beyond 2~kpc and reach further into the halo.

Internal extinction may bias the estimated disc rotation to slightly lower values at the midplane than is true for the ionised gas. Measuring the rotation exactly at the midplane (rather than using our peak-velocity extraction) would result in a $30\rm\ km\ s^{-1}$ increase in the final outflow velocity profile. We explicitly account for this phase-specific offset with out approach to measure $\upsilon_{\rm rot}$ mentioned above, to ensure that the kinematic profiles presented in Fig.~\ref{fig:velprofs} accurately reflect the true velocity difference between the ionised gas in the outflow and the one in the disc.

We evaluated the impact of using different kinematic estimators, specifically the moment analysis used for CO versus Gaussian line-fitting for H$\alpha$. To ensure consistency, we convolved the ALMA data to the MUSE spectral resolution and compared the resulting line shapes. We found that the emission lines remain largely symmetric and that the final velocity profiles and outflow properties do not vary significantly for the cold gas if the convolved cube is used instead of the original cube. This cross-phase consistency suggests that beam-smearing and line-of-sight velocity gradients do not significantly bias our comparison between the molecular and ionised outflow components.

\subsection{Effects of rotational lag in the extraplanar gas}\label{sec:velprofs_wlag}

If outflows entrain material from a rotating disc, then some circular motion remains as it moves outward. Extraplanar emission in both neutral and ionised phases is widely observed to co-rotate with the underlying disc, though more slowly, exhibiting a characteristic rotation lag \citep[e.g.,][]{Fraternali2002,Cooper2008,Marasco2019, Li2021}. The preserved rotation supports a scenario where the expelled material originates from dense disc clouds that are ablated and entrained into the fractal medium of the wind \citep{Greve2004, Cooper2009}.

The definition described in Sect.~\ref{sec:velprofs_nolag} assumes that vertical motion dominates the kinematic residual between the extraplanar gas and the disc. Here, we investigate possible signatures of rotation in the extraplanar material and test how our deprojected velocity profiles vary when part of the velocity is attributed to the expected rotation of the gas as it moves into the CGM.

To model this rotational lag, we extracted rotation curves from horizontal slices across the minor axis at increasing distances from the midplane. We estimate the amount of rotation in each slice by measuring the difference in velocity at the edge of the rotation curve (on the major-axis) from the velocity directly above the galaxy centre $\delta V_{\rm lag}(z)\equiv V(z,r) - V(0,r)$. The observed rotation lag decreases with distance from the midplane for both molecular and ionised phases. We find that $\delta V_{\rm lag}(z)$ is well approximated by a $z^{-1}$ decay, mirroring the rotation parameter ($\gamma$) of \citet[see their Fig. 7]{Greve2004}. We therefore fit a relationship of $\delta V_{\rm lag}(z)\propto z^{-1}$ to our data and remove the rotational lag from our \vlos.

For the molecular phase, we find a rotational lag of $60\rm\ km\ s^{-1}$ at $1\rm\ kpc$, dropping to $30\rm\ km\ s^{-1}$ at $2\rm\ kpc$. This becomes effectively negligible ($\delta V_{\rm lag}(z)\sim0\rm\ km\ s^{-1}$) at $2.5\rm\ kpc$. The ionised phase exhibits a slightly slower rotation of $50\rm\ km\ s^{-1}$ at $1\rm\ kpc$, reaching $0\rm\ km\ s^{-1}$ at $3\rm\ kpc$. This reduction in rotational lag over $\sim$2~kpc effectively steepens the vertical velocity gradient. 

This reduction is consistent with HI observations of edge-on systems, where the rotational lag is often larger closer to the midplane before flattening further out \citep{Oosterloo2007, Marasco2019}. These works find gradients of -40 to -50~$\rm km\ s^{-1}\ kpc^{-1}$. The average extraplanar rotational lag observed across larger samples of nearby spirals typically ranges from -10 to -20~$\rm km\ s^{-1}\ kpc^{-1}$ \citep{Li2021}. Our values are therefore within the range of previous work. 

Under this assumption, within the disc ($z\lesssim\pm0.5$~kpc), rotational motion dominates the kinematics, beyond which the motion perpendicular to the disc becomes more prominent as the rotational component decreases. Consistent with \cite{Kim2018}, we observe that the rotation becomes negligible at higher altitudes, a transition theoretically attributed to the exchange of angular momentum with the CGM \citep{Fraternali2008,Fraternali2017, Marasco2019}. This interaction stalls the rotation of the outflowing gas, causing a loss of centrifugal support that allows vertical and inward radial motions to dominate the kinematics beyond the disc-halo interface.

Compared to the profiles in Fig.~\ref{fig:velprofs}, the rotation-compensated velocities exhibit a similar shape, but are systematically slower. The median difference is $30\rm\ km\ s^{-1}$, with peak deviations of $95\rm\ km\ s^{-1}$ for the ionised gas and $60\rm\ km\ s^{-1}$ for the molecular phase. Both models converge at the furthest detected distances from the midplane, confirming that the impact of rotation is confined primarily to the outflow base interface.

We note that apart from a region near the outflow base ($z\sim\pm1$~kpc), the rotation-compensated velocities are fully enclosed in our uncertainties presented in Fig.~\ref{fig:velprofs}. We discuss how this modified kinematics affects the outflow physical properties in Sect.~\ref{sec:whatifrotation}.

\section{Outflow properties in ESO~484-036}\label{sec:outrates}

Having established the outflow nature of the extraplanar emission through its morphology and kinematics, we now calculate the most relevant physical properties of the molecular and ionised phases in the outflow.

\subsection{Molecular mass from CO}\label{sect:comass}

Molecular mass (\mmol) for the entire galaxy and the outflow region can be calculated using the following equation:
\begin{equation}
    M_{\rm mol} = \alpha_{\rm CO} \cdot L'_{\rm CO}\ [\rm \rm M_\odot]\\
\end{equation}
with $\alpha_{\rm CO}$ being the CO-to-H$_2$ conversion factor in units of $\ \rm M_\odot (\rm K\ km\ s^{-1}\ pc^2)^{-1}$ and $L'_{\rm CO}$ the luminosity of CO in units of $\rm K\ km\ s^{-1}\ pc^2$.

Although $\alpha_{\rm CO}$ is often considered a significant source of uncertainty, the fact that ESO~484-036 is a metal-rich starburst system suggests a fairly low $\alpha_{\rm CO}$. In \cite{Bolatto2013review}, the authors review the conditions for $\alpha_{\rm CO}$, and we expect the range of likely values to be 0.8$-$1.5~$\rm M_\odot (\rm K\ km\ s^{-1}\ pc^2)^{-1}$, and therefore should not be considered to increase the systematic uncertainty by much more than $30\%$.

The gas excitation, opacity, and abundance ratios between CO/H$_2$ are known to be different in outflow regions, and it also varies from case to case \citep{Bolatto2013, Leroy2015, Veilleux2020, Pereira2024}. Our choice of $\alpha_{\rm CO}=1$ is primarily driven by the strong obscuration in both the disc and the dusty outflow (based on detections of PAH emission in NIRCam/MIRI imaging). This justifies adopting an $\alpha_{\rm CO}$ value higher than the optically thin limit of $\alpha_{\rm CO}=0.34$ for the outflow, and comparable to that used for the starbursting disc. The use of a single value of $\alpha_{\rm CO}$ is common in the literature and eases comparison with other results \citep[e.g.,][]{Leroy2015, Krieger2019, Lopez2020}.

We calculate a log(\mmol/M$_\odot$)~$=9.13$ for the entire galaxy and log(\mmol/M$_\odot$)~$=8.42$ for the outflow region, representing $19\%$ of the total \mmol. These values are also presented in Table~\ref{tab:outflow}. 

\begin{table*}
    \centering
    \caption{Outflow properties derived for each phase observed in ESO~484-036}
    \label{tab:outflow}
    \renewcommand{\arraystretch}{1.3}
    \setlength{\tabcolsep}{0pt}
    \begin{tabular*}{\textwidth}{@{\extracolsep{\fill}}lcccccccc}
        \toprule
        \multicolumn{1}{l}{Outflow Phase} & 
        \multicolumn{1}{c}{$M_{\rm out}$} &
        \multicolumn{1}{c}{$M_{\rm out}$/$M_{\rm total}$} &
        \multicolumn{1}{c}{$\dot{M}$} & 
        \multicolumn{1}{c}{$\eta_{M}$} & 
        \multicolumn{1}{c}{$\dot E$} & 
        \multicolumn{1}{c}{$\eta_{E}$} &
        \multicolumn{1}{c}{$\dot p$} &
        \multicolumn{1}{c}{$\eta_p$} \\
        & \multicolumn{1}{c}{$\log\left({M_{\rm out}}/{\rm M_\odot}\right)$} &
        \multicolumn{1}{c}{\%} &
        \multicolumn{1}{c}{M$_\odot$ yr$^{-1}$} &
        \multicolumn{1}{c}{$\dot{M}/\rm SFR$} &
        \multicolumn{1}{c}{$\log\left(\dot{E}/{\rm erg\ s^{-1}}\right)$} &
        \multicolumn{1}{c}{$\dot{E}/\dot E_*$} &
        \multicolumn{1}{c}{$\log\left(\dot{p}/\mathrm{dynes}\right)$} &
        \multicolumn{1}{c}{$\dot{p}/\dot p_*$} \\
        (1) & (2) & (3) & (4) & (5) & (6) & (7) & (8) & (9) \\
        \midrule
        Molecular (CO) &
        $8.42$ &
        $19.3$ &
        $13.02-53.74$ &
        $1.50-6.18$ &
        $40.00-41.73$ &
        $0.003-0.14$ &
        $33.58-34.76$ &
        $0.10-1.43$ \\
        Ionised (H$\alpha$) &
        $7.24$ &
        $16.3$ &
        $1.24-4.81$ &
        $0.14-0.55$ &
        $39.46-40.95$ &
        $0.001-0.02$ &
        $32.81-33.85$ &
        $0.02-0.18$ \\
        \bottomrule
    \end{tabular*}
    \flushleft\footnotesize{\textbf{Notes.} Col. (1): Phase of the gas associated with the outflow; Col. (2): Total mass of the outflow; Col. (3): Ratio of outflow mass to the total gas mass; Col. (4): Mass outflow rate; Col. (5): Mass loading factor; Col. (6): Energy flux; Col. (7): Energy loading factor; Col. (8): Momentum flux; Col. (9): Momentum loading factor. The ranges correspond to the values derived assuming half-opening angles ($\theta$) between $22^\circ-40^\circ$.}
\end{table*}

\subsection{Ionised mass from H$\alpha$}\label{sec:ionisedmass}

\begin{equation}
    M_{\rm ion} = 3.3\times10^8\frac{C_e\cdot L_{\rm H\alpha,44}}{n_{e,3}}\ [\rm \rm M_\odot]
\end{equation}
with $C_e\equiv\langle n_e^2\rangle/n_e^2$ the clumpiness factor, $L_{\rm H\alpha,44}$ the luminosity of H$\alpha$ normalised to $10^{44}\ \rm erg\ s^{-1}$ and $n_{e,3}$ the average electron density ($n_e$) normalised to $10^3\ \rm cm^{-3}$ for the ionised mass. 

Clumpiness introduces uncertainty without clear constraints. For uniform-density clouds, $C_e \sim 1$, while stronger density variations in clumpier gas produce larger $C_e$. Simulations suggest that values above unity are realistic \citep[e.g.][]{Schneider2018}. Because we lack any observational constraints, we set $C_e = 1$. This is common in these cases \citep{Veilleux2020,Mazzilli2025}.

We acknowledge that the behaviour of $n_e$ is much more complex than a fixed value \citep[e.g.,][]{Huang2026}. The work of \cite{Mazzilli2025} shows that for NGC~4666, $n_e$ decreases until $\sim2$~kpc from the midplane, with an increase to $200-300\rm\ cm^{-3}$, then constant values for a larger distance. The work of \cite{Xu2023} shows that for M82, $n_e$ decreases to 2~kpc from the midplane, but they do not have data beyond this. Unresolved measurements find high $n_e$ in the outflows of ULIRGs \citep{Fluetsch2019}.

Strong skyline contamination within the [SII] doublet makes a direct measurement of a $n_e$ profile within the outflow regions impossible. The disc is the only location where the [SII] emission is sufficiently robust to allow for a measurement. We measure $n_e$ values between $80-110\rm\ cm^{-3}$. Due to the observational limitations in the extraplanar gas, we adopt a single, uniform value of $n_e = 100\rm\ cm^{-3}$ for the entire system (both disc and outflow), following the work by \cite{Mazzilli2025}.

In order to calculate the \mion\ of the galaxy and outflow, we use the IR/Optical curve of \cite{Calzetti2000} to calculate the extinction correction for each pixel using the Balmer-decrement, H$\alpha$/H$\beta$, a $R_V = 3.1$ and a $[\rm H\alpha/H\beta]_{\rm int}=2.86$, typical for a temperature $T_e = 10^4$ K \citep{Osterbrock2006} and standard for star-forming galaxies in the literature \citep[][]{Calzetti2000, Calzetti2001, Draine2021}.

The H$\beta$ map contains fewer valid pixels than the H$\alpha$ map when a $\rm S/N = 5$ is used in both cases. To ensure a complete correction across the full extent of the outflow, instead of a pixel-by-pixel approach, we use an extinction profile, $\rm E(B-V)(z)$, by calculating the median Balmer decrement as a function of distance from the midplane along the minor axis.

This profile captures the dominant vertical gradient in the dust column while allowing us to apply a consistent correction to regions where H$\beta$ is not detected. The resulting extinction profile peaks at the galactic centre with a median $\rm E(B-V) \sim 1.4$ (reaching a maximum of $\sim 2$ in individual pixels). Applying the adopted extinction profile gives a central $A_V \sim 4.3$ and $A_{\rm H\alpha} \sim 3.5$.

Despite using the profile for global correction, a pixel-by-pixel analysis remains essential for interpreting the central morphology. We find that the bright emission-line region located directly west (right) of the nucleus in the midplane (Fig.~\ref{fig:Halpha_maps}, top panel) is not a physical local maximum. When the pixel-by-pixel correction is applied, this feature disappears, confirming it is an artifact of nuclear obscuration. The apparent peak simply identifies the location where the dust column is thin enough for H$\alpha$ photons to escape, while the true intrinsic peak at the nucleus remains suppressed until corrected.

Once the H$\alpha$ map is extinction corrected, we calculate a log(\mion/M$_\odot$)~$=8.02$ for the entire galaxy and log(\mion/M$_\odot$)~$=7.24$ for the outflow region, representing $16\%$ of the total \mion. These values are also presented in Table~\ref{tab:outflow}. Note that this is an order of magnitude less massive than the \mmol\ present in the same region.

\subsection{Multiphase mass outflow rates}

\begin{figure}
    \centering
    \includegraphics[width=0.5\textwidth]{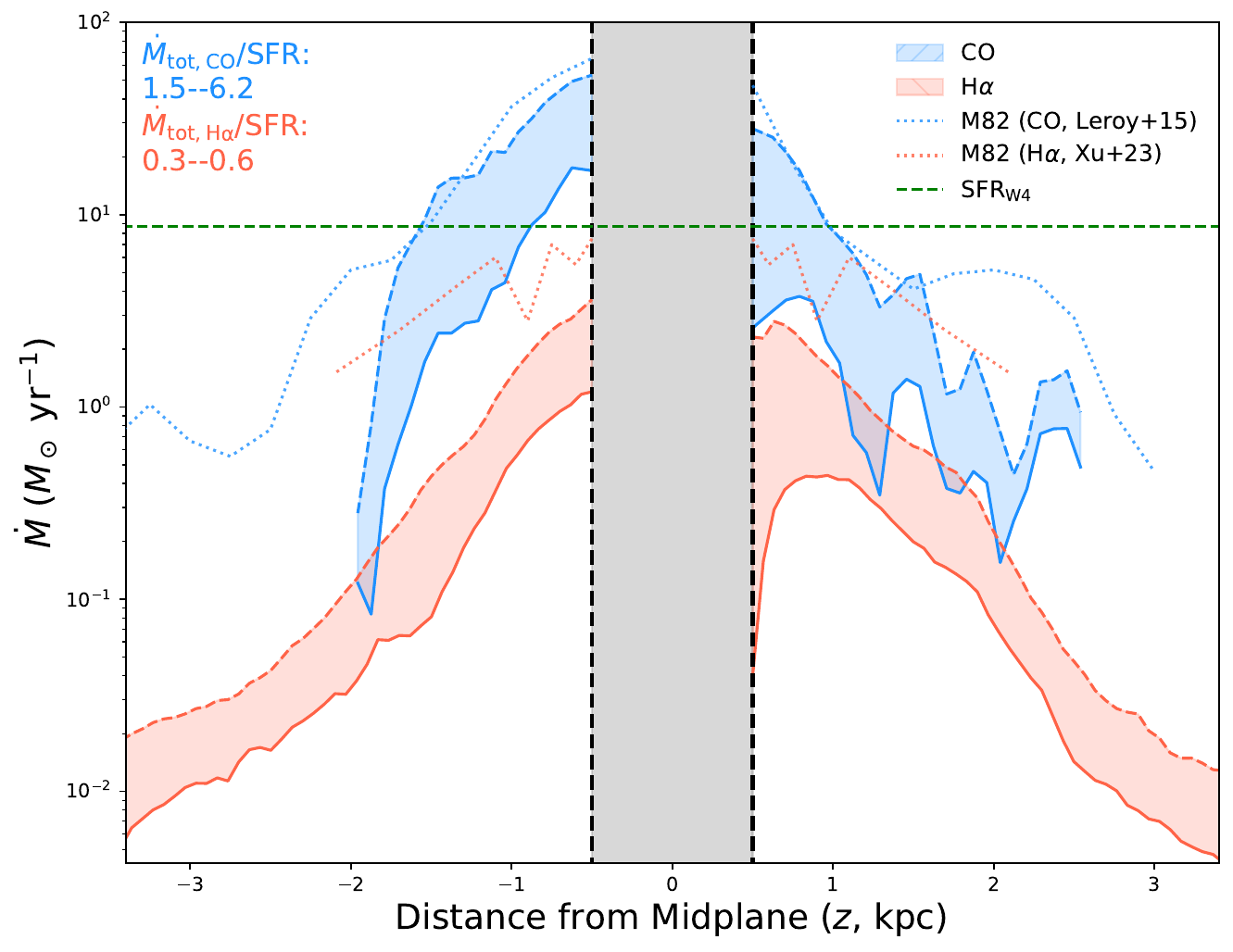}
    \caption{Mass outflow rate (\mdot) profiles for ESO~484-036 for each phase. Coloured shaded regions represent the effect of varying the half-opening angle ($\theta$) between $22^\circ-40^\circ$. Horizontal green line represents the $\rm SFR=8.7\ \rm M_\odot\ yr^{-1}$. Grey band is the disc region. For comparison, we plot \mdot\ profiles from M82 based on CO \protect\citep{Leroy2015} and H$\alpha$ \protect\citep{Xu2023} in dotted lines. The latter is mirrored, due to it being available only for the south of M82.}
    \label{fig:M_Profiles}
\end{figure}

To obtain a mass outflow rate (\mdot) profile and total value, we use the following formulas:
\begin{equation}
\begin{split}
    \dot{M}(z) &= \frac{M(z)\cdot \upsilon_{\rm out}(z)}{\Delta z}\ [\rm \rm M_\odot\ yr^{-1}]\\
    \dot M_{\rm tot} &= \sum^{z_{\rm max}}_{z=z_{\rm base}} \frac{M(z)\cdot \upsilon_{\rm out}(z)}{z} [\rm \rm M_\odot\ yr^{-1}]
\end{split}
\end{equation}
For the \mdot\ profile, $M(z)$ is the mass at a given $z$, $\Delta z$ is the physical scale along the minor axis of the galaxy in the observations (MUSE is 66 pc per pixel and ALMA is 83 pc per pixel) and $\upsilon_{\rm out}(z)$ is the non-rotational-lag velocity profile calculated in Sect.~\ref{sec:velprofs_nolag} (see Sect.~\ref{sec:whatifrotation} for effects of rotation lag in outflow properties). Note that $M(z)/\Delta z$ is equivalent to the linear mass density in \cite{Leroy2015}. For the total \mdot, we use $z_{\rm base}=0.5$ kpc, being the base of the outflow defined by the end of the disc region in Fig.~\ref{fig:RGB}. This differs from adding values in the profile, as $z$ is not a fixed value. 

Our profiles for \mdot\ are presented in Fig.~\ref{fig:M_Profiles}. Both molecular and ionised phases have declining profiles, indicating that mass loss is efficient near the launching site, compared with larger distances. Since the velocity profiles present positive gradients with distance or are flat, it is specifically the mass per element that is decreasing. The molecular gas in the south may still exist but is so diffuse that ALMA cannot detect it, which explains the abrupt decline in the molecular \mdot\ profile. The upper \mdot\ profile for the molecular phase has similar shape and values as M82 \citep{Leroy2015}, but declines much faster between 1.5 and 2 kpc, reaching \mdot\ similar to the ionised phase at this distance on both sides of the disc.

The ionised gas \mdot\ profile extends up to 3.5~kpc on both sides of the midplane. Near the launching site, the ionised \mdot\ profile lies about 1~dex below the molecular \mdot\ profile near the base of the outflow. Due to the interaction between phases, we expect mass injection from the molecular phase to the ionised phase \citep{Kim2020,Schinnerer2024}. This interaction can explain the non-detection of molecular gas beyond 2~kpc, because the gas becomes too diffuse and faint.

The total \mdot\ for the cold phase is between $13-54\ \rm M_\odot\ \rm yr^{-1}$, comparable to other starburst driven outflows in the literature \citep[e.g.,][]{Leroy2015, Fluetsch2019}, although we highlight that this range of values may be affected by deprojection issues. This gives a mass loading factor ($\eta_M=\dot M/\rm SFR$) between $1.5-6.2$ for the cold phase, ejecting substantial amounts of gas that will not be available for star formation. For the ionised phase, \mdot\ is between $1-5\ \rm M_\odot\ \rm yr^{-1}$, also comparable to other starburst driven outflows in the literature \citep{Xu2022}, giving a $\eta_M$ between $0.1-0.6$ for this phase. These values are also presented in Table~\ref{tab:outflow}

\subsection{Mass removal efficiency and depletion time}

Overall, the molecular phase clearly dominates the mass budget of the outflow, with a mass outflow rate that is higher by $\sim 1$~dex compared to the ionised phase, being consistent with other purely SF-driven outflows \citep{Bolatto2013, Leroy2015, Mazzilli2025}. The multiphase mass loading factor ($\eta_{M,\rm cold} + \eta_{M,\rm ion}$) shows that ESO~484-036 actively removes gas at a rate approximately seven times higher than its SFR, highlighting an efficient mass-loss mechanism.

The rapid mass-removal rate observed in ESO~484-036 has the potential to quickly deplete the molecular gas reservoir and quench active star formation. To quantify this, we calculate the total depletion timescale ($\tau_{\rm dep}$) by dividing the molecular gas mass in the disc by the sum of the SFR and the multiphase mass outflow rate ($\dot{M}_{\rm out}$). We derived $\tau_{\rm dep} = 16-48\rm\ Myr$. It is significantly shorter than the typical $\sim 100\rm\ Myr$ seen in average starbursts \citep{Bolatto2024}, aligning instead with the high-SFR regions of extreme systems like IRAS 08339+6517 \citep{ReichardtChu2022} and approaching, though not reaching, the rapid $1-10\rm\ Myr$ timescales of AGN-driven outflows \citep{Cicone2014}. 

However, the long-term impact of this feedback depends on the ultimate fate of the gas. As noted in Sect. \ref{sec:dynamics}, the deprojected velocities in the southern outflow ($\lesssim400\rm\ km\ s^{-1}$) remain largely below $\upsilon_{\rm esc}\sim550\rm\ km\ s^{-1}$. Given the outflow velocities, a recycling flow is a plausible scenario; in this case, part of the entrained material will eventually cool and fall back onto the disc. Whether the gas is photoionised by the CGM or shock-heated during its transit remains an open question. Such thermal processing could delay the cooling of the recycled gas, potentially inhibiting star formation when the material eventually falls back into the disc. Although the outflow clears the disc at rates approaching AGN-driven systems, this feedback likely results in regulation of star formation rather than a complete, immediate quenching.

\subsection{Energy and momentum flux in the outflow}

\cite{Thompson2024} review the mechanisms that drive outflows, showing how each mechanism injects a distinct amount of energy and momentum into a wind. From this, we can estimate the outflows kinetic energy and momentum and directly compare them to the energy output predicted by a given launch mechanism. Any mismatch between the two then reveals which physical model we must invoke to explain the outflow. For example, if star formation alone, primarily through SNe, does not provide enough energy/momentum, we may need to attribute part of the driving to the AGN \citep[e.g.,][]{Cicone2014, Chisholm2017}.

\begin{figure}
    \centering
    \includegraphics[width=0.5\textwidth]{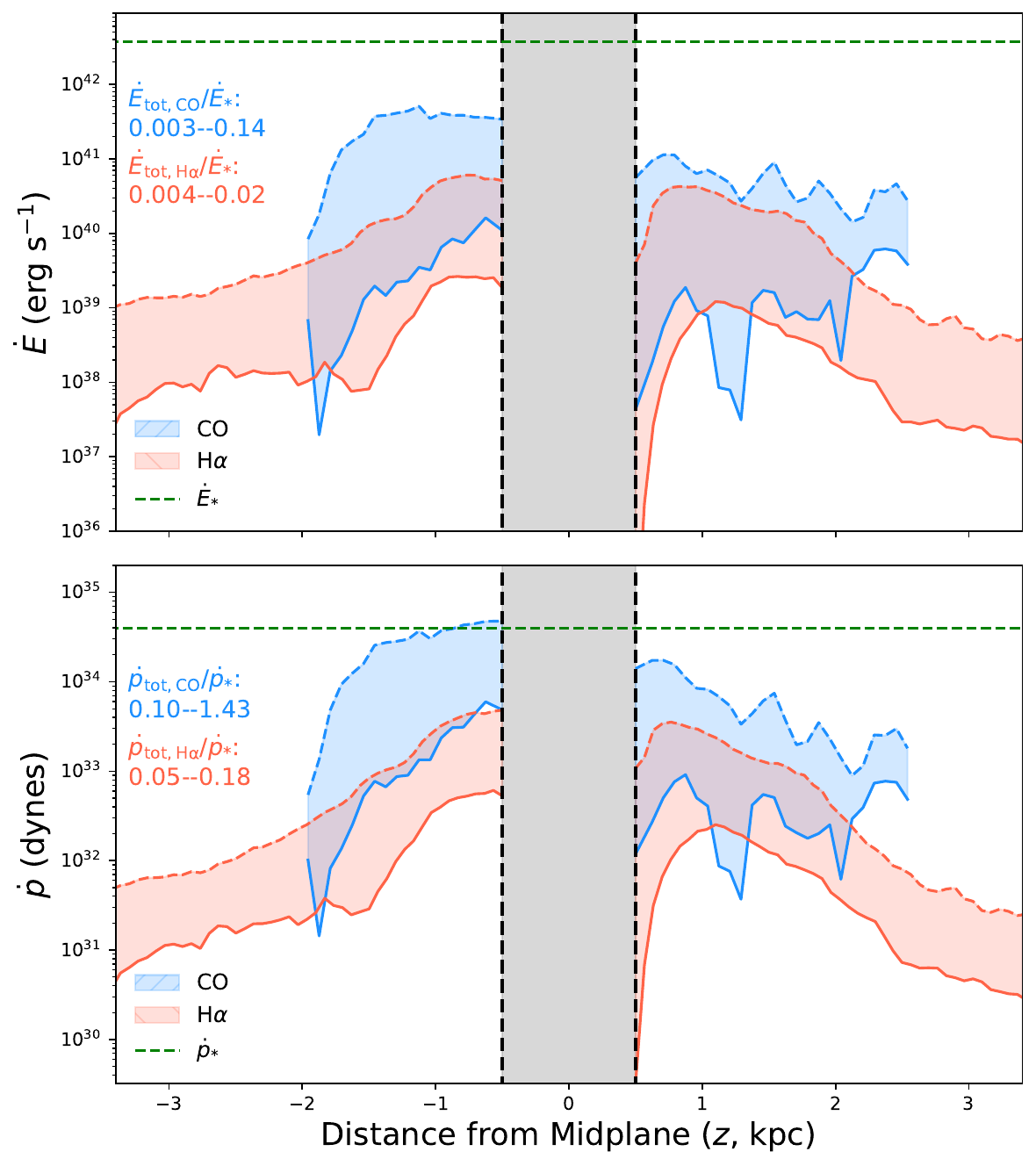}
    \caption{Energy flux (top) and momentum flux (bottom) profiles for ESO~484-036 for each phase. Coloured shaded regions represent the effect of varying the half-opening angle ($\theta$) between $22^\circ-40^\circ$. Horizontal green line represents the energy and momentum generated from SF. Grey band is the disc region.
    Combined energy loading factor of both phases is high ($\eta_E\lesssim0.2$) but consistent with a starburst-driven wind without AGN input. The starburst alone supplies enough momentum flux ($\eta_p\lesssim1$) to drive the outflow without any additional energy source.}
    \label{fig:E_P_Profiles}
\end{figure}

Kinetic energy and momentum flux (\edot\ and \pdot) are intrinsically related to the driving mechanism of the outflow, and can help determine the true nature of it. In order to obtain profiles for \edot\ and \pdot\ we use the following formulas:
\begin{equation}
\begin{split}
    \dot{E}(z) &= \frac{1}{2}\dot M(z) \cdot \left( \upsilon_{\rm out}(z)\right)^2\ [\rm erg\ s^{-1}]\\
    \dot{p}(z) &= \dot M(z) \cdot \upsilon_{\rm out}(z)\ [\rm dynes]\\
\end{split}
\end{equation}
and for total \edot\ and \pdot\ we use the following formulas:
\begin{equation}
\begin{split}
    \dot E_{\rm tot} &= \sum^{z_{\rm max}}_{z=z_{\rm base}}\frac{M(z)\cdot \left(\upsilon_{\rm out}(z)\right)^3}{2\cdot z}\ [\rm erg\ s^{-1}]\\
    \dot p_{\rm tot} &= \sum^{z_{\rm max}}_{z=z_{\rm base}}\frac{M(z)\cdot \left(\upsilon_{\rm out}(z)\right)^2}{z}\ [\rm dynes]
\end{split}
\end{equation}
Similarly to \mdot, this differs from summing values in the profile, as $z$ is not a fixed value. 

We calculate the energy and momentum injected by SNe produced in the starburst using the formulas from \cite{Xu2022}:
\begin{equation}
\begin{split}
    \dot{E}_{*} &= 4.3\times10^{41}\rm\times SFR\ [erg\ s^{-1}]\\
    \dot{p}_{*} &= 4.6\times10^{33}\rm\times SFR\ [dynes]
\end{split}
\end{equation}
with SFR in $\rm \rm M_\odot\ yr^{-1}$. For ESO~484-036 this gives a $\log(\dot E_*/\rm erg\ s^{-1}) = 42.57$ and $\log(\dot p_*/\rm dynes) = 34.6$ that consider only the contribution of SNe energy and momentum injection and no other contributions as stellar winds or cosmic rays.

We obtain the energy loading factor ($\eta_E = \dot E/\dot E_*$) and the momentum loading factor ($\eta_p = \dot p/\dot p_*$), which provide far more information than the energy and momentum flux alone. All energy and momentum values are also presented in Table~\ref{tab:outflow}

Our profiles for \edot\ and \pdot\ are presented in Fig.~\ref{fig:E_P_Profiles}. These represent our most uncertain estimates because the systematic uncertainty is significantly amplified by the velocity dependence of these properties ($\dot{p} \propto v^2$ and $\dot{E} \propto v^3$). 
These values should be treated with caution, as the true energetics probably lie near a median value.

The \edot\ profile on both sides of the midplane appears relatively flat in the cold phase.
Near 2~kpc, the south outflow drastically decreases, dropping by almost 2~dex in the lower profile. The north profile varies significantly because the velocity profile strongly affects it, with differences reaching up to 1~dex.

The \edot\ profile of ionised gas is quite symmetric, decreasing first then flattening beyond $\sim1.5$~kpc in the south and $\sim2$~kpc in the north. Although the south profile is systematically lower than the cold gas profile, north profile shows similar values to the cold phase. Despite the slightly higher velocities in the ionised phase, the cold gas carries an order of magnitude more mass; this translates to a significantly higher \edot, confirming that the molecular phase carries more energy in the outflow.

The cold outflow is highly efficient, with a $\log(\dot E/\rm erg\ s^{-1})$ between 40 and 41.7, thus obtaining an energy loading $\eta_E\lesssim0.14$. This value lies at the lower end of the measured energy coupling in the SF-driven outflow in NGC~4945 \citep[0.15 to 0.4,][]{Bolatto2021}, and it agrees well with the even lower energy loadings reported in other cases where SF alone can explain the energetics \citep[0.01 to 0.1,][]{Cicone2014, Krieger2019}.

The driving of the ionised phase is also efficient, with $\log(\dot E/\rm erg\ s^{-1})$ between 39.46 and 41, thus an energy loading $\eta_E\lesssim0.02$. This value is in agreement with the efficiencies expected for SF-driven outflows \citep[0.02 to 0.21,][]{Chisholm2017,Cronin2025}.

The total coupling efficiency for ESO~484-036 is $\eta_E\lesssim0.2$. This lies within the expected range of SF-driven outflows and agrees with the expected theoretical value of $\eta_E \sim 0.1$ \citep{Thompson2024}, where SNe and stellar winds can explain the energetics in the outflow. Even accounting for deprojection uncertainties, our measurements indicate that SF alone can drive the outflow in ESO~484-036, without any need for AGN contribution.

The common criterion for knowing whether it is possible for a starburst to drive the outflow using momentum injection is $\eta_p\lesssim1$ \citep{Hopkins2012,Chisholm2017,Cronin2025}. In that case, momentum imparted by the starburst (force) is enough to drive the outflow without the need for any other driver, such as an AGN. The ionised phase shows \etap\ between $0.02-0.18$, with a complete range of values that meet the criteria. The cold phase \etap\ is between $0.1-1.43$, also consistent with an SF-driven outflow, considering the uncertainties. 

The combined total \etap\ is between $0.12-1.61$. Our high value for the total \etap\ exceeds unity. This can be explained by the hot gas pressure and the radiation pressure of the dust \citep{Chisholm2017}. The high extinctions of $\rm E(B-V)\sim2$ at the centre of the galaxy can imply a mechanism in which momentum from the hot fluid is transferred to cold neutral gas. This would mean that stellar feedback remains a plausible explanation for this pressure \citep{Lopez2011,Lopez2014}. We therefore do not need to invoke an AGN. Note that the momentum injection only considers SNe as a source of momentum, so the inclusion of stellar winds should only increase this injection, thus making $\eta_p$ smaller \citep{Thompson2024}.

\subsection{Impact of rotational lag on estimates of outflow mass loss and energetics}\label{sec:whatifrotation}

The mass outflow rate, momentum flux, and energy flux scale with velocity ($\dot{M} \propto v, \dot{p} \propto v^2, \dot{E} \propto v^3$). Accounting for rotational lag will therefore produce changes in the estimates of these values. For the molecular phase, the rotation-compensated \mdot\ decreases to $41.37\rm\ M_\odot\ yr^{-1}$ (a reduction of $23\%$), resulting in a mass-loading factor of $\eta_M \approx 4.8$. The impact on the ionised phase is more significant, with the corrected \mdot\ dropping by $\sim 60\%$ to $1.91\rm\ M_\odot\ yr^{-1}$ ($\eta_M \approx 0.22$).

Molecular gas dominates both the energy and momentum loading of the outflow in ESO~484-036. This implies that the correction for rotational lag in the ionised gas has a smaller impact on the total energy and momentum loading. For the molecular phase, the rotation-corrected momentum loading remains significant at $\eta_p \approx 1.05$, while the energy loading remains at $\eta_E \approx 0.11$. In contrast, the ionised phase contributes negligibly to the total energetics under these assumptions ($\eta_p \approx 0.06, \eta_E \approx 0.008$).

Although the rotation-compensated values are systematically lower, the fundamental interpretation of the ESO~484-036 outflow as a molecular-dominated starburst-driven event remains unchanged. Incorporating rotational lag adds a necessary layer of physical rigour and underscores that velocity remains the primary source of systematic uncertainty in characterising these winds. However, our core results remain robust, the mass-loading and energetics of the outflow are easily explained with a starburst, comparable with starburst-driven systems in the literature.

\subsection{Towards a systematic ratio of ionised-to-molecular gas mass loading}

\begin{figure}
    \centering
    \includegraphics[width=0.5\textwidth]{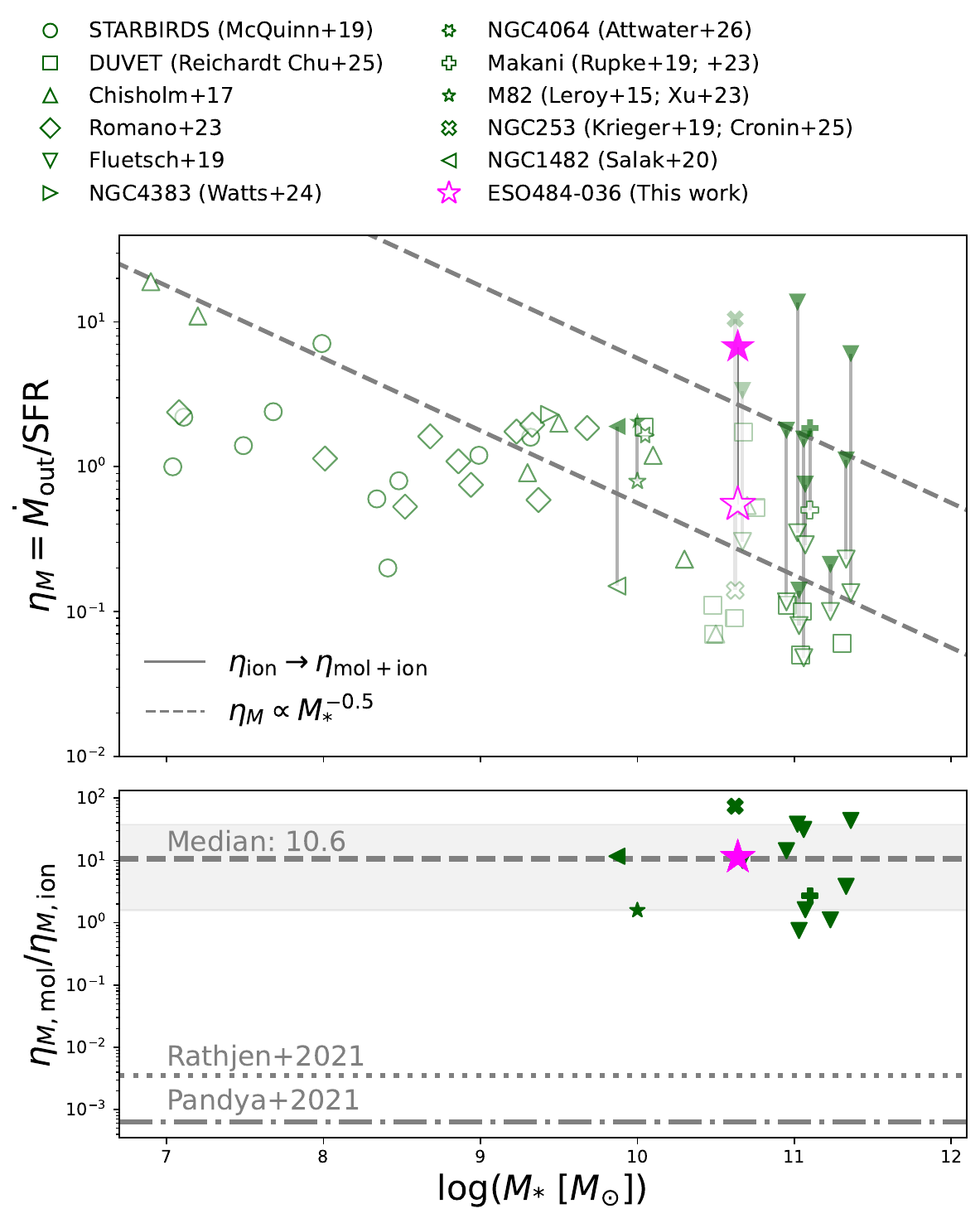}
    \caption{Top: Mass loading factor ($\eta_M$) against stellar mass (\mstar) for ESO~484-036 and a sample of galaxies in the Local Universe available in the literature. Empty symbols show single-phase (ionised or neutral atomic) measurements, while filled symbols denote multiphase (ionised and molecular gas) measurements. Vertical grey lines strictly connect measurements between single-phase ionised gas to multiphase measurements including both ionised and molecular gas. ESO~484-036 is highlighted in magenta. Literature values with similar stellar masses to ESO~484-036 are more transparent for visualisation. Dashed grey lines show theoretical scaling relations with a slope matching the FIRE-2 simulation. The lower line is normalised to the data median, and the upper line is shifted 1~dex higher. Bottom: Ratio between molecular and ionised mass loading factors ($\eta_M$) against stellar mass (\mstar), only showing galaxies with both measurements available. Median ratio is presented with a dashed grey line with $\pm1\sigma$ band around it. For comparison, we also present the ratios derived from the results of \protect\cite{Rathjen2021} and \protect\cite{Pandya2021}, highlighting the discrepancy.
    }
    \label{fig:eta_v_mass}
\end{figure}

In Fig.~\ref{fig:eta_v_mass}, we compare mass loading factors (\etam) to host galaxy stellar mass. We do this for ESO~484-036 and compare to objects in the literature. Note that we use single-phase $\eta_M$ for NGC~4383 \citep{Watts2024}, NGC~4064 \citep{Attwater2026} and from multiple compilations \citep{Chisholm2017, McQuinn2019, Romano2023, ReichardtChu25}. In the case of mutiphase $\eta_M$, including both ionised and molecular phase, we use the resolved studies of Makani \citep{Rupke2019, Rupke2023}, M82 \citep{Leroy2015, Xu2023}, NGC~253 \citep{Krieger2019, Cronin2025}, NGC~1482 \citep{Salak2020}, and the sample of sources from \cite{Fluetsch2019}.

We highlight ESO~484-036 in magenta, showing the maximum \etam\ for our ranges. Note that these values can be smaller, but the ratio between phases stays constant. Physically, we expect \etam\ to decrease as \mstar\ increases \citep{Pandya2021}. More massive galaxies possess deeper gravitational potential wells, which require significantly higher energy and momentum injection to expel gas into the CGM.

We compare our results with theoretical scaling relations using a fixed slope of $-0.5$ (Fig.~\ref{fig:eta_v_mass}, top panel). This slope matches the predictions of the FIRE-2 simulation \citep{Pandya2021} and is similar to the relation $\eta_M \propto M_*^{-0.4}$ derived from ionised gas absorption lines by \cite{Chisholm2017}. 

In the figure, we show $\eta = a\times M_*^{-0.5}$ with two values of the scale-factor ($a$). The first (lower line) is set to match the median of all ionised gas values, which we find to be $\log(a)\sim4.7$. The second (top line) is shifted 1~dex higher, showing good alignment with multiphase values combining ionised and molecular gas. This offset illustrates that, although the slope of the scaling relation may be similar across phases, the total mass-loss rate is much higher once the molecular, dense ISM is included.

The ionised \etam\ roughly follows the median scaling across the mass range of $\log(M_*/\rm M_\odot) \sim 9-11$. However, at lower masses, the relation $\eta \propto M_*^{-0.5}$ can overestimate the mass loading of the ionised gas, when $\log(M_*/\rm M_\odot) \lesssim 8.5$. There are no molecular phase estimates (using CO) of mass loading at low masses. We therefore cannot determine whether this is true for the total mass loading. 

In our sample, the molecular-to-ionised mass-loading ratio ($\eta_{\rm mol} / \eta_{\rm ion}$) shows a median value of $10.6$, although with significant scatter spanning $\pm0.8$~dex (Fig. \ref{fig:eta_v_mass}, bottom panel). Although this distribution excludes a universal scaling law, it demonstrates a consistent trend in which the molecular phase dominates the mass budget throughout the sampled mass range, being different to the ratios found in previous theoretical work \citep{Rathjen2021, Pandya2021}. The significant scatter in this ratio likely stems from systematic uncertainties in the mass derivations, primarily the choice of $\alpha_{\rm CO}$ and $n_e$ (Sects. \ref{sect:comass} and \ref{sec:ionisedmass}). Although the observations of molecular phase outflows are biased towards higher masses, $\log(M_*/\rm M_\odot) = 10.5-11.5$, we do observe that high molecular-to-ionised ratios persists in systems with masses as low as $\sim10^{10}$~M$_{\odot}$, like M82 and NGC~1482. This indicates that the dominance of the molecular phase appears to be a fundamental characteristic of these starburst-driven winds over all ranges currently observed.

Our results in Fig.~\ref{fig:eta_v_mass} highlight a discrepancy between the observed molecular outflows and the theoretical work. The cosmological simulation FIRE-2 \citep{Pandya2021} finds that the molecular-to-total \etam\ ratio is negligible ($\sim 10^{-3.2}$) for local galaxies ($z < 0.5$), suggesting that the molecular phase contributes less than $1\%$ to the mass budget of the outflow. We note that there are very significant differences in the way outflow mass loading is measured in observations and simulations \citep[for discussion see][]{ReichardtChu2022}. In the work of \cite{Pandya2021}, the limit where the extraplanar material is considered an outflow is defined as $0.1-0.2\ R_{\rm vir}$ (roughly 25~kpc for a Milky Way-like halo). In contrast, CO emission in ESO~484-036 is detected only up to 2.5~kpc from the disc, and even H$\alpha$ emission does not reach these distances, with detection stopping at 4~kpc for $\rm S/N=5$.

Higher resolution magneto-hydrodynamic simulations from \cite{Rathjen2021} also predict low fractions of molecular gas mass-loss ($\lesssim 0.3\%$) using the SILCC ISM box simulations. \cite{Kim2018} finds a similar low ratio of molecular gas mass compared to ionised gas using the TIGRESS simulations. The \cite{Rathjen2021} result is lower than our observed median ratio of $\eta_{M,\rm mol}/\eta_{M,\rm ion} \sim 11$ by 3.5~dex. Moreover, both \cite{Kim2018} and \cite{Rathjen2021} find essentially no molecular phase gas travelling beyond $\sim1$~kpc from the galaxy midplane, which is different from our results. This discrepancy indicates that current simulations do not reproduce the production and/or survival of molecular gas in starburst-driven winds.

We note that the results of \cite{Kim2020} appear consistent with our measured ratios in the lower panel of Fig.~\ref{fig:eta_v_mass}. However, their definition of cool gas combines phases with $T\sim10^4\rm\ K$, and is therefore not directly comparable to our molecular-to-ionised ratios. As the authors note, depending on the distance from the midplane, their definition of cool gas is effectively the same as the warm gas used in \cite{Kim2018}.
  
Our results indicate that the molecular gas is the dominant component in comparison to the ionised phase. We do not know the mass loading in the hot, X-ray-emitting gas in the wind. Simulations suggest that this is an important mass component of the wind \citep{Rathjen2021, Pandya2021}. Moreover, instantaneous measures of \etam\ struggle to disentangle the time-delays between peak star formation and subsequent outflow propagation predicted by models \citep{Pandya2021}. This time-delay should be assuaged by measuring a larger number of targets. Reconciling these differences requires a combination of high-resolution simulations, alongside multiphase observations that span from the molecular to the hot ionised regimes on larger numbers of galaxies.

\section{Summary}\label{sec:conclusion}

In this work, we present spatially resolved observations of the ionised gas using H$\alpha$ and molecular gas using CO(1$-$0), in the nearby starburst galaxy ESO~484-036. These data provide the first complete multiphase characterisation of the outflow in this system. Our main findings are as follows.

\begin{enumerate}
\item Using VLT/MUSE and ALMA observations from the GECKOS survey, we resolved the ionised and molecular gas. These data provide a rare, high-resolution view of the spatial and kinematic interactions between gas phases in the extraplanar emission present in this edge-on system. The molecular gas extends to $\sim 2.5$~kpc, surrounding a central ionised core that reaches beyond $\sim$4~kpc.

\item We resolved spatially co-located extraplanar CO(1$-$0) and H$\alpha$ emission tracing a truncated biconical (frustum) outflow ($i \sim 87.5^\circ$, $\theta \sim 22^\circ-40^\circ$). At the outflow base ($0.5-1.0 \text{ kpc}$), the structure consists of a centrally peaked ionised core surrounded by a molecular sheath. Robust molecular detection and heavy midplane obscuration indicate a significant dust reservoir entrained in this powerful, starburst-driven wind.

\item Multiphase kinematics shows both components keeping some rotation from the underlying disc, appearing blueshifted to the north and redshifted to the south. Anti-correlated velocity dispersions, centrally peaked $\sigma_{\rm H\alpha}$ and V-shaped $\sigma_{\rm CO}$, support a scenario where the fast ionised wind efficiently entrains and transfers momentum to the surrounding molecular sheath.

\item The multiphase kinematics reveal a complex interplay between the gas components. Although the ionised gas in the south exhibits a positive gradient from $\sim 180\rm\ km\ s^{-1}$ to $\sim 280\rm\ km\ s^{-1}$ at $3\rm\ kpc$, we observe a convergence between the ionised and molecular gas velocities within the $1-2\rm\ kpc$ range. Despite this localised alignment, a significant north-south kinematic asymmetry persists in the cold phase. The northern molecular outflow reaches mean velocities of $\sim 270\rm\ km\ s^{-1}$, whereas the southern component maintains a lower mean of $\sim 150\rm\ km\ s^{-1}$.

\item Outflow velocities remain well below the escape velocity ($\sim 550\text{ km s}^{-1}$). A plausible scenario is a large-scale recycling flow, where a fraction of the expelled gas eventually stalls and returns to the galactic disc.

\item The molecular phase dominates the mass budget, with a mass outflow rate ($\dot{M}_{\rm mol} \sim 13-54\ \rm \rm M_\odot\ yr^{-1}$) roughly 1~dex higher than the ionised rate ($\dot{M}_{\rm ion} \sim 1-5\ \rm \rm M_\odot\ yr^{-1}$). The high total mass loading factor ($\eta_{M} \sim 7$) implies that the galaxy is efficiently losing mass, giving a depletion time ($\tau_{\rm dep}$) between $16-48\ \rm Myr$. The outflow may regulate rather than permanently quench the gas reservoir given the measured outflow velocities.

\item The loading factors for energy ($\eta_E \le 0.2$) and momentum ($\eta_p \le 1$) are entirely consistent with stellar feedback. No AGN contribution is required to drive the observed energetics, even considering extreme estimates of the systematic uncertainty.

\item The high molecular mass loading factor tracks a scaling relation 1~dex higher than single-phase ionised studies. This highlights a fundamental discrepancy with cosmological simulations, which underestimate the cold-to-total mass ratio by $\sim 3.5$~dex. This tension likely arises because simulations often use outflow boundaries (e.g., $R_{\rm vir}$) that exclude short-range outflows or are unable to produce cold gas in the sub-parsec scales required. Our results demonstrate that these recycling flows are the primary mass-carriers in starburst-driven winds, even if they do not escape the galactic halo.

\end{enumerate}

\section*{Acknowledgements}
Based on observations made with ESO Telescopes at the
La Silla Paranal Observatory under program ID 110.24AS. We wish to thank the ESO staff, and in particular the staff at Paranal Observatory, for carrying out the GECKOS observations. This paper makes use of the following ALMA data: ADS/JAO.ALMA\#2023.1.00698.S. ALMA is a partnership of ESO (representing its member states), NSF (USA) and NINS (Japan), together with NRC (Canada), NSTC and ASIAA (Taiwan), and KASI (Republic of Korea), in cooperation with the Republic of Chile. The Joint ALMA Observatory is operated by ESO, AUI/NRAO and NAOJ. 
This work is based, in part, on observations made with the NASA/ESA/CSA James Webb Space Telescope. The data were obtained from the Mikulski Archive for Space Telescopes at the Space Telescope Science Institute, which is operated by the Association of Universities for Research in Astronomy, Inc., under NASA contract NAS 5-03127 for JWST. These observations are associated with program \#5637.
This paper makes use of services that have been provided by AAO Data Central (datacentral.org.au).
Part of this research was conducted by the Australian Research Council Centre of Excellence for All Sky Astrophysics in 3 Dimensions (ASTRO 3D), through project number CE170100013.

MM acknowledges support from the UK Science and Technology Facilities Council through grant ST/Y002490/1.

LC acknowledges support from the Australian Research Council
Discovery Project funding scheme (DP210100337).

LMV acknowledges support by the German Academic Scholarship Foundation (Studienstiftung des deutschen Volkes) and the Marianne-Plehn-Program of the Elite Network of Bavaria.

 This research used the following software: Astropy \citep{astropy}, Numpy \citep{numpy}, Scipy \citep{scipy}, Matplotlib \citep{matplotlib} and Multicolorfits \citep{multicolorfits}.

\section*{Data Availability}

The GECKOS survey is still in progress, but the VLT/MUSE ESO~484-036 data used in this work is available in the archive (program: 110.24AS). 
Both the ALMA/ACA (project: 2023.1.00698.S) and the JWST (GO project: 5637) observations are publicly available.




\bibliographystyle{mnras}
\bibliography{main.bib} 

\appendix
\section{Impact of Inclination on Outflow Properties}\label{app:variations}

As discussed in Sect.~\ref{sec:geometry}, we obtained a stellar disc inclination of $i=84^\circ$ from JWST F150W and F200W imaging. However, we adopted the GECKOS survey value of $i = 87.5^\circ$ for our analysis. In this appendix, we quantify how adopting the lower inclination ($i = 84^\circ$) affects our results. A comparison of the derived properties is presented in Table~\ref{tab:outflow_comparison}.

\begin{table}
    \centering
    \caption{Outflow properties comparison by inclination}
    \label{tab:outflow_comparison}
    \begin{tabular*}{\columnwidth}{@{\extracolsep{\fill}}lc}
        \toprule
        Outflow Property & Value Range / Unit \\
        \midrule
        \multicolumn{2}{c}{Cold Gas} \\
        \midrule
        Mass outflow rate ($\dot{M}$) & $M_{\odot}$ yr$^{-1}$\\
        \hspace{1em} with $i = 87.5^{\circ}$ & $13.02 - 53.74$ \\
        \hspace{1em} with $i = 84^{\circ}$   & $13.73 - 59.19$ \\
        \addlinespace
        Mass loading factor ($\eta_M$) & $\dot M/\mathrm{SFR}$\\
        \hspace{1em} with $i = 87.5^{\circ}$ & $1.50 - 6.18$ \\
        \hspace{1em} with $i = 84^{\circ}$   & $1.58 - 6.80$ \\
        \addlinespace
        Energy flux ($\dot{E}$) & $\log\left(\dot{E}/\mathrm{erg\ s^{-1}}\right)$\\
        \hspace{1em} with $i = 87.5^{\circ}$ & $40.00 - 41.73$ \\
        \hspace{1em} with $i = 84^{\circ}$   & $40.08 - 41.92$ \\
        \addlinespace
        Energy loading factor ($\eta_E$) & $\dot E/\dot E_*$\\
        \hspace{1em} with $i = 87.5^{\circ}$ & $0.003 - 0.14$ \\
        \hspace{1em} with $i = 84^{\circ}$   & $0.003 - 0.22$ \\
        \addlinespace
        Momentum flux ($\dot{p}$) & $\log\left(\dot{p}/\mathrm{dynes}\right)$\\
        \hspace{1em} with $i = 87.5^{\circ}$ & $33.58 - 34.76$ \\
        \hspace{1em} with $i = 84^{\circ}$   & $33.64 - 34.87$ \\
        \addlinespace
        Momentum loading factor ($\eta_p$) & $\dot p/\dot p_*$\\
        \hspace{1em} with $i = 87.5^{\circ}$ & $0.10 - 1.43$ \\
        \hspace{1em} with $i = 84^{\circ}$   & $0.11 - 1.85$ \\
        
        \midrule
        \multicolumn{2}{c}{Ionised Gas} \\
        \midrule
        Mass outflow rate ($\dot{M}$) & $M_{\odot}$ yr$^{-1}$\\
        \hspace{1em} with $i = 87.5^{\circ}$ & $1.24 - 4.81$ \\
        \hspace{1em} with $i = 84^{\circ}$   & $1.28 - 5.02$ \\
        \addlinespace
        Mass loading factor ($\eta_M$) & $\dot M/\mathrm{SFR}$\\
        \hspace{1em} with $i = 87.5^{\circ}$ & $0.14 - 0.55$ \\
        \hspace{1em} with $i = 84^{\circ}$   & $0.15 - 0.58$ \\
        \addlinespace
        Energy flux ($\dot{E}$) & $\log\left(\dot{E}/\mathrm{erg\ s^{-1}}\right)$\\
        \hspace{1em} with $i = 87.5^{\circ}$ & $39.46 - 40.95$ \\
        \hspace{1em} with $i = 84^{\circ}$   & $39.53 - 41.07$ \\
        \addlinespace
        Energy loading factor ($\eta_E$) & $\dot E/\dot E_*$\\
        \hspace{1em} with $i = 87.5^{\circ}$ & $0.001 - 0.02$ \\
        \hspace{1em} with $i = 84^{\circ}$   & $0.001 - 0.03$ \\
        \addlinespace
        Momentum flux ($\dot{p}$) & $\log\left(\dot{p}/\mathrm{dynes}\right)$\\
        \hspace{1em} with $i = 87.5^{\circ}$ & $32.81 - 33.85$ \\
        \hspace{1em} with $i = 84^{\circ}$   & $32.85 - 33.91$ \\
        \addlinespace
        Momentum loading factor ($\eta_p$) & $\dot p/\dot p_*$\\
        \hspace{1em} with $i = 87.5^{\circ}$ & $0.02 - 0.18$ \\
        \hspace{1em} with $i = 84^{\circ}$   & $0.02 - 0.20$ \\
        \bottomrule
    \end{tabular*}
\end{table}

\begin{figure}
    \centering
    \includegraphics[width=0.5\textwidth]{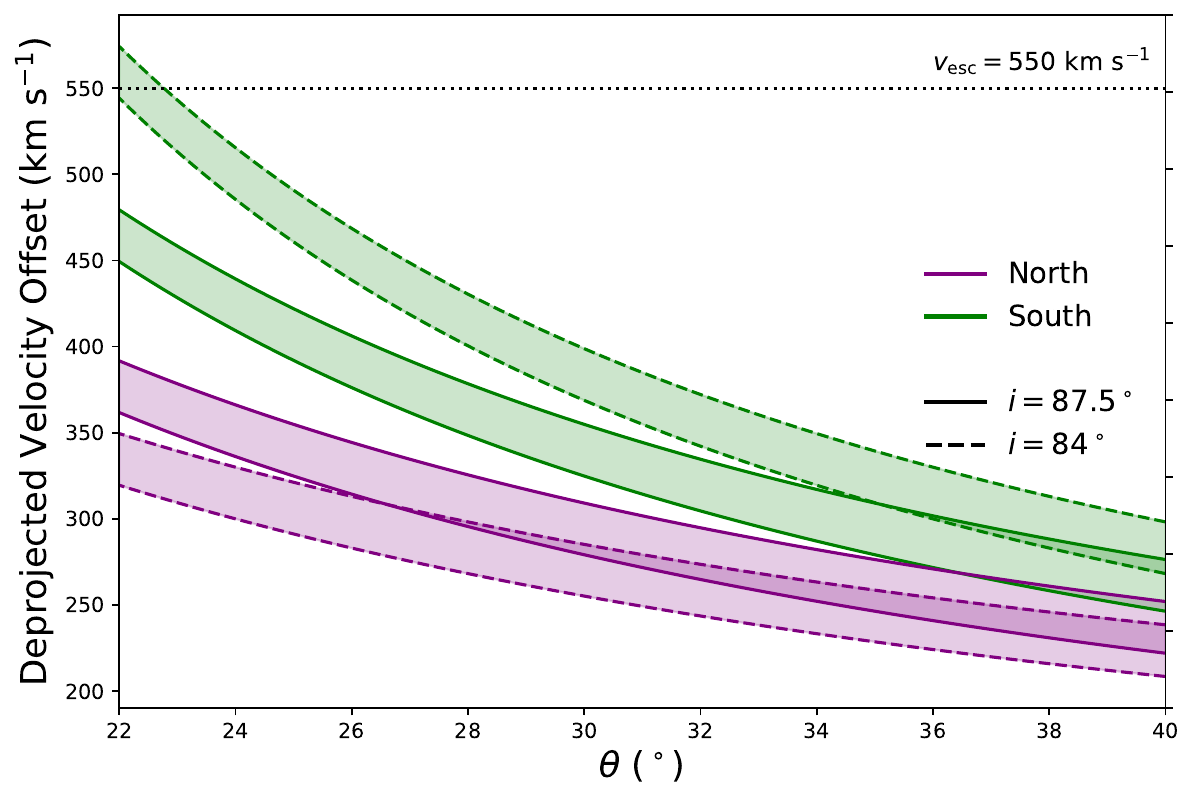}
    \caption{Geometric stress test of the deprojected velocity ($\upsilon_{\rm out}$) as a function of the half-opening angle ($\theta$). The curves represent the deprojection of the peak observed velocity offset ($150\ \rm km\ s^{-1}$) in the ionised gas for two different inclinations: $i = 87.5^\circ$ (solid lines) and $i = 84^\circ$ (dashed lines). The purple curves correspond to the north (approaching) outflow, while the green curves correspond to the south (receding) outflow. Shaded regions show the $30\ \rm km\ s^{-1}$ increase produced if the rotation is measured exactly at the midplane. The grey horizontal line marks the escape velocity ($\upsilon_{\rm esc} \approx 550\ \rm km\ s^{-1}$). Even at the most extreme geometric limits (narrow $\theta$ and lower $i$), the deprojected velocities barely exceeds $\upsilon_{\rm esc}$, remaining largely sub-escape for all possible $\theta$.}
    \label{fig:velvariations}
\end{figure}

Although the assumed biconical geometry remains the same, with the half-opening angle ($\theta$) between $22^\circ-40^\circ$, the choice of inclination ($i$) significantly impacts the deprojection factor, $\alpha = |\cos(i \pm \theta)|$. Based on the \vlos\ maps of CO and H$\alpha$, we define the north outflow as approaching the observer (blueshifted) and the south outflow as receding (redshifted). 

In Fig.~\ref{fig:velvariations}, we show the sensitivity of the deprojected velocity ($\upsilon_{\rm out}$) to different assumed $i$. To provide a stress test for our geometric model, we deproject the ionised peak velocity offset of $150\ \rm km\ s^{-1}$. Although this peak is localised to only a few pixels in the north (approaching) ionised outflow, it serves as a benchmark for assessing how systematic uncertainties in $i$ and $\theta$ propagate to our final kinematic estimates.

For the north (approaching) cone, a lower inclination ($i = 84^\circ$) increases the deprojection factor $\alpha$. For example, with a narrow half-opening angle ($\theta = 22^\circ$), $\alpha$ increases from $0.414$ ($i = 87.5^\circ$) to $0.469$ ($i = 84^\circ$). This decreases $\upsilon_{\rm out}$ from $385\ \rm km\ s^{-1}$ down to $340\ \rm km\ s^{-1}$.

For the south (receding) cone, the effect is more dramatic. As the inclination decreases, the cone axis moves further from the line-of-sight. For $\theta = 22^\circ$, $\alpha$ drops to $0.275$ at $i = 84^\circ$. This deprojects the peak velocity offset up to $\sim 580\ \rm km\ s^{-1}$. Although this specific high velocity likely represents a geometric overestimation due to the narrow $\theta$, even with a median $\theta = 30^\circ$, the lower inclination yields a velocity of $\sim 400\rm \ km\ s^{-1}$, compared to $\sim 340\rm \ km\ s^{-1}$ at $i = 87.5^\circ$. These variations become negligible when the broadest half-opening angle is assumed ($\theta = 40^\circ$). 

For the cold molecular gas, decreasing the inclination from $i=87.5^\circ$ to $i=84^\circ$ produces minor changes to the derived outflow properties. The mass outflow rate ($\dot{M}$) range shifts slightly from $13.0-53.7$ to $13.7-59.2\ \rm \rm M_\odot\ yr^{-1}$, while the mass loading factor ($\eta_M$) remains effectively consistent, peaking at $6.8$ (Table \ref{tab:outflow_comparison}). Similarly, the energy and momentum budgets show only marginal increases; the maximum energy loading factor ($\eta_E$) moves from $0.14$ to $0.22$, and the momentum loading factor ($\eta_p$) from $1.43$ to $1.85$. Overall, the cold gas results remain robust against these variations in inclination. 

The ionised gas properties exhibit similar variations. Although the lower values for all properties remain stable, the high value of the mass outflow rate increases from $4.81$ to $5.02\ \rm \rm M_\odot\ yr^{-1}$ assuming $i=84^\circ$. The change is negligible in the momentum and energy. The maximum \etae\ shifts from $0.02$ to $0.03$, and \etap\ increases from $0.18$ to $0.20$. Consequently, while a lower inclination implies a slightly more energetic ionised phase, the total mass budget remains comparable across both assumptions. 

Regardless of the inclination assumed, the resulting differences are small relative to the dominant systematic uncertainties of the system. The reported for $\dot{E}$ and $\dot{p}$ are highly sensitive to the deprojection of velocities that are nearly perpendicular to the line-of-sight in this edge-on orientation. These limits likely represent conservative low and high values, and we emphasise that the true physical properties of the outflow in ESO~484-036 likely reside closer to the median values of our reported ranges.

\section{Constraining the outflow driver using BPT diagnostics}

\begin{figure}
    \centering
    \includegraphics[width=0.5\textwidth]{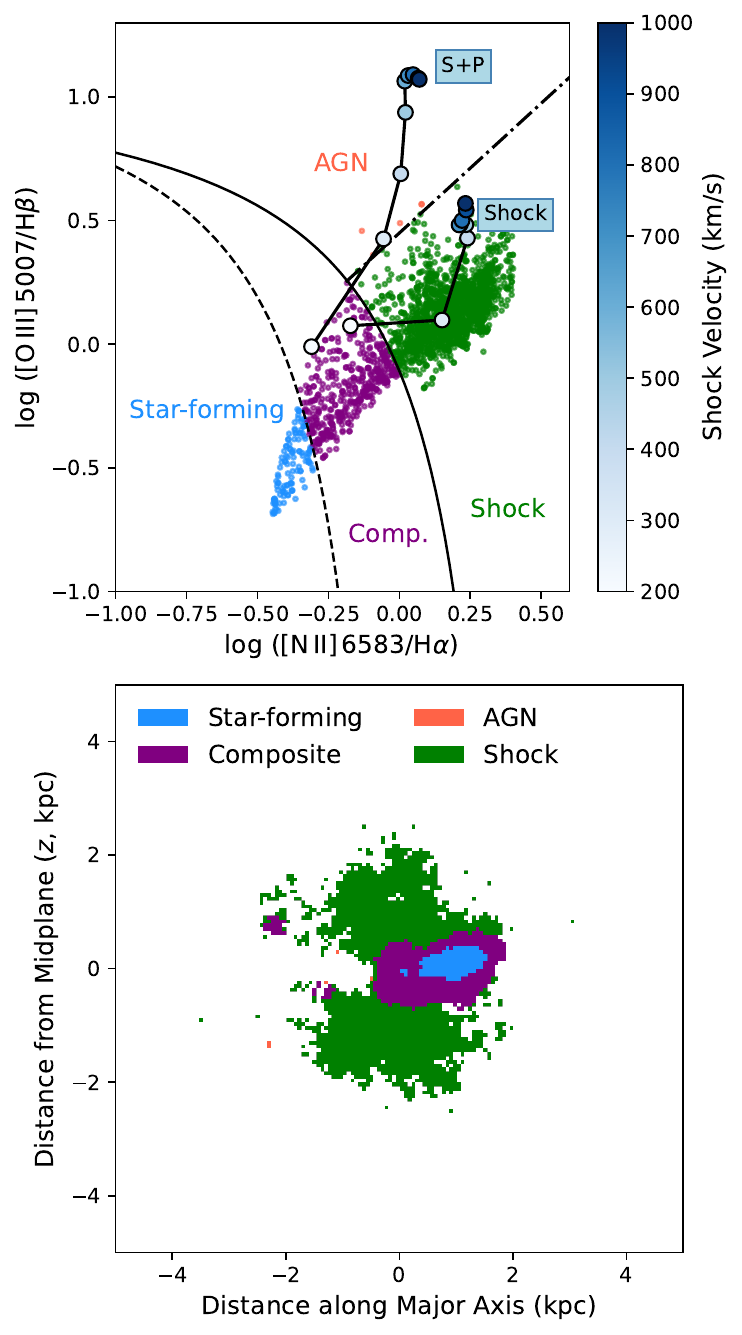}
    \caption{Diagnostic emission-line ratios and spatial ionisation structure of ESO~484-036. Top: Pixel-by-pixel BPT diagram ($\log(\rm[OIII]/H\beta)$ vs. $\log(\rm[NII]/H\alpha)$) color-coded by diagnostic region: star-forming (blue), composite (purple), shock (green), and AGN (red). Overlaid black tracks represent shock and shock+precursor (S+P) models from \protect\cite{Allen2008}, color-coded by shock velocity (right color bar). Bottom: Spatial distribution of the corresponding ionisation classes relative to the galactic midplane and minor axis. The northern and southern outflow sides exhibit ionisation consistent with stellar shocks of $\sim300\rm\ km\  s^{-1}$ without a precursor. The bulb of star-forming emission west of the nucleus requires cautious interpretation. Although it shows an $E(B-V) \sim 1$, it remains significantly affected by the high midplane obscuration. Extreme central extinction ($E(B-V) \sim 2$) and the non-detection of H$\beta$ ($\rm S/N=5$) limits further interpretation of the sparse pixels consistent with AGN ionisation.}
    \label{fig:bpt}
\end{figure}

The nature of the central engine driving the outflow in ESO~484-036 is a subject of ongoing investigation. Although global classifications for this galaxy vary, ranging from a pure starburst to obscured AGN (see Sect.~\ref{sec:target}), the high central obscuration and nearly edge-on orientation ($i \sim 87.5^{\circ}$) significantly limit the effectiveness of traditional emission-line diagnostics.

The dusty disc produces extreme extinction at the galactic centre ($\rm E(B-V) \sim 2$). In this edge-on configuration, masking the nucleus, preventing a direct view of the central ionising source. Traditional BPT diagnostics (e.g., $\rm[OIII]/H\beta$ versus $\rm[NII]/H\alpha$) are highly susceptible to being dominated by surface emission rather than the true nuclear signature.

Figure~\ref{fig:bpt} (top panel) displays the pixel-by-pixel BPT diagram for ESO~484-036, mapping ionisation variations across the distinct regions shown in the bottom panel. This resolved approach allows us to separate the star-forming disc from the shock-dominated extraplanar gas. 

Ionisation in both the northern and southern outflow sides aligns with stellar shock models of $\sim300\rm\ km\ s^{-1}$ without a precursor. Although the emission bulb west of the nucleus suggests pure SF ionisation, this feature requires cautious interpretation. Even with an extinction of $\rm E(B-V)\sim1$ (half the value measured in the nucleus), this region remains significantly affected by dust. Extreme obscuration hides the true nucleus and eastern disc, where H$\beta$ remains undetected ($\rm S/N=5$).

Although sparse pixels appear consistent with AGN ionisation, the non-detection of H$\beta$ in the surrounding eastern disc limits further interpretation. 

This diagnosis remains inconclusive. Although a hidden AGN could contribute to the galactic wind, our energetic and momentum analysis suggests that the starburst alone provides sufficient power to drive the observed outflow without requiring an additional energy source.








\bsp	
\label{lastpage}
\end{document}